\documentclass{revtex4}

\usepackage{graphicx}
\usepackage{amsmath}

\usepackage{color}

\def\giorno{14/7/2015}

\def\a{\alpha}
\def\b{\beta}

\def\ga{\gamma}
\def\de{\delta}   
\def\eps{\varepsilon}
\def\phi{\varphi}
\def\la{\lambda}

\def\om{\omega}

\def\L{\mathcal{L}}

\def\P{\mathcal{P}}

\def\R{{\bf R}}

\def\pa{\partial}

\def\xb{{\bf x}}
\def\yb{{\bf y}}
\def\hb{{\bf h}}

\def\o+{\oplus}

\def\<{\langle}
\def\>{\rangle}

\def\Tr{\mathtt{Tr}}

\def\({\left(}
\def\){\right)}
\def\[{\left[}
\def\]{\right]}
\def\=#1{\bar #1}
\def\~#1{\widetilde #1}
\def\wt#1{\widetilde #1}
\def\.#1{\dot #1}
\def\^#1{\widehat #1}
\def\"#1{\ddot #1}

\def\eeq{\end{equation}}
\def\beq{\begin{equation}}

\def\beql#1{\begin{equation} \label{#1}}

\def\eqref#1{(\ref{#1})}

\def\Fb{{\bf F}}

\begin{document}

\title{Poincar\'e-like approach to Landau theory. \\
II. Simplyfying the Landau-deGennes potential for nematic liquid crystals}

\author{Giuseppe Gaeta}
\email{giuseppe.gaeta@unimi.it}
\affiliation{Dipartimento di Matematica, Universit\`a degli Studi di Milano, via Saldini 50,
I-20133 Milano (Italy)}

\date{\giorno}

\begin{abstract}
In a previous paper we have discussed how the Landau potential
(entering in Landau theory of phase transitions) can be simplified
using the Poincar\'e normalization procedure. Here we apply this
approach to the Landau-deGennes functional for the
isotropic-nematic transitions, and transitions between different
nematic phases, in liquid crystals. {We give special attention to
applying our method in the region near the main transition point,
showing in full detail how this can be done via a suitable simple
modification of our Poincar\'e-like method. We also consider the
question if biaxial phases can branch directly off the fully
symmetric state; some partial results in this direction are
presented.}
\end{abstract}



\maketitle

\section*{Introduction.}

In the Landau theory of phase transitions \cite{L1,Lan5} the state
of a system is described by minima of a group-invariant potential
$\Phi (x)$, depending on the order parameters $x$ and on external
physical parameters (temperature,  pressure,...) which control the
phase transition. As these are varied, the minima of the Landau
potential change location and assume different invariance
properties, so corresponding to different phases.

The Landau potential is a polynomial one, and the order \footnote{By
this we mean of course the degree of the polynomial; it is customary
to use the word ``order'' in this context since the degree of the Landau polynomial
corresponds to the order in perturbation theory we are considering.} $N$ of
this is sometimes not so low; moreover all possible terms (that
is, invariant monomials) of order up to $N$ should be included.
Thus in concrete applications the Landau potential can be rather
complex, and the study of its minima depending on parameters
can result quite difficult.

It would of course be convenient to have a criterion for
simplifying the Landau potential, i.e. to be able to drop certain
of the invariant monomials. Different criteria to this effect have
been proposed in the literature, and among these we mention in
particular the one by Gufan \cite{Guf,SGU}, then recast in an
earlier paper by the present author \cite{AOP}.

In the companion paper \cite{GL1} we have provided a simple
discussion of how the technique created by Poincar\'e in the
framework of dynamical systems to simplify nonlinear terms
\cite{ArnG,Elp,Wal,CGs,CiW,Wency,Gae02,RNF} could be adapted to
the framework of Landau theory; note that in this context one
should most often refrain from a complete simplification of the
highest order terms, as the requirement of thermodynamic stability
-- that is, convexity at large distances from the origin -- should
be taken into account, as discussed in \cite{GL1} and recalled
here in Sect.\ref{sec:convex}.

This Poincar\'e-like approach is completely algorithmic and quite
general; in particular, it can be pushed up to any desired order
(albeit some convergence issues should be controlled; these are
related to small denominators and will be completely explicit in
any concrete computation; see the remarks at the end of
Sect.\ref{sec:red6}). \footnote{It should also be recalled that
Poincar\'e-Birkhoff normal forms have been used to compute quantum
spectra with a remarkable level of accuracy up to near
dissociation threshold, see e.g. \cite{Joy,JoyS}.}

In the present note we want to proceed along the same lines to
analyze a concrete problem, i.e. the Landau-deGennes (LdG in the
following) theory for the isotropic-nematic transitions in liquid
crystals \cite{deG,Vir,GLJ,AlL}; this also describes transitions
between different nematic phases (biaxial, and different uniaxial
ones, in particular oblate and prolate uniaxial). {In this
respect, we will devote special attention to the possible direct
transition between the isotropic and a biaxial phase \cite{Fre}.

More generally, we consider how our method can be applied near the
main transition point; at this point the linear (homological)
operator on which the Poincar\'e procedure is based is degenerate,
so that the standard method does not apply; however, our general
method \cite{GL1} is formulated in such a way to be readily
applicable in this more general case. We show in full detail how
this is done operationally; this implicitly makes use of the
``further normalization'' technique developed for dynamical
systems \cite{Gae02,RNF}, but we will not need to discuss it
here.}

It should be stressed that in this note, at difference with the
general discussion given in \cite{GL1}, we do not aim at
mathematical generality, but have a very concrete case -- i.e. a
given Landau potential, the LdG one -- at hand. Also we know the
order of the LdG potential. This means that some of the issues in
the general approach proposed in \cite{GL1} can be substantially
streamlined (see Appendix B for a comparison); in particular, we
do not need to proceed order by order but can proceed by a direct
computation \footnote{Which is implemented by an algebraic
manipulation language, given the considerable complexity of the
algebra involved.} taking into account the main ideas behind the
Poincar\'e approach.

As well known, in the LdG theory \cite{deG,Vir,GLJ,AlL} the order
parameter is a tensorial one (this corresponds to the general
situation in liquid crystals \cite{Vir}), i.e. a three-dimensional
symmetric traceless matrix $Q$; and the symmetry group, which in
this case is the three-dimensional rotation group $SO(3)$, acts on
it by conjugations, i.e. $Q \to R Q R^{-1}$. \footnote{Note the
effective order parameters reduce to the two independent
eigenvalues $\la_1$, $\la_2$ of $Q$, as different $Q$'s related by
a similarity transformation are equivalent.}
\bigskip

The {\it plan of the paper} is as follows. We will first analyze
in details this $SO(3)$ action (Sect.\ref{sec:so3}) and the LdG
functional (Sect.\ref{sec:LdG}); and then pass to implement the
Poincar\'e approach on the LdG functional. We will first report on
the results of computations up to order six (Sect.\ref{sec:red6}),
as it would follow from the fact the two basic invariants are of
order two and three (see Sects.\ref{sec:so3} and \ref{sec:LdG});
but in subsequent analysis it will result that working at this
order we find a non-physical degeneration (which also leads to
non-physical results), so that we will extend our computations and
analysis to order eight (Sect.\ref{sec:red8}); this will eliminate
the degeneracy met at order six, and give agreement with classical
results \cite{AlL}. Sect.\ref{sec:unibireg} is devoted to the
analysis of the reduced LdG obtained through our method. Here we
will find out that, as mentioned above, the computations at order
six yield a non-physical degeneration, which in turn gives results
in disagreement with the widely accepted ones \cite{AlL}. On the
other hand, going at the next order in our expansion, i.e. order
eight (a functional of order seven is forbidden by the request of
thermodynamic stability) we remove the degeneracy, and it turns
out the results obtained via the reduced functional are then in
agreement with those of the full theory \cite{AlL}. {These
computations are performed under a non degeneracy assumption for
the quadratic term of the LdG potential, which fails precisely at
the main transition point; so they do not apply in the vicinity of
the main phase transition. This region of the parameter space is
studied in Sect.\ref{sec:unibising}; here we will in particular
look for biaxial solutions, and conclude that under some
non-degeneracy assumptions (involving only coefficients of higher
order terms in the LdG potential) the branches identified in
Sect.\ref{sec:unibireg} are unstable at the main bifurcation.} We
conclude our work by summarizing and discussing our results in
Sect.\ref{sec:conc}. Two brief Appendices are devoted to the
Molien function (App.A), and to comparison with previous work
(App.B). {Two other Appendices collect some involved formulas
related to the discussion in Sect.\ref{sec:red8} (App.C) and in
Sect.\ref{sec:SP8} (App. D).}
\bigskip

We would like to mention that in our computations {several of }
the ``intermediate'' terms can be eliminated from the LdG
potential by a suitable change of coordinates \footnote{It should
be noted this will be non-homogeneous; thus the disappearing terms
are actually just recombined to a simpler expression in the new
coordinates}; the relevant point here is that the change of
coordinates needed to reach this simpler form can, and will, be
{\it explicitly} computed.

We also mention that our work here follows the same approach as in
the companion paper \cite{GL1}; however, there we considered only
effects of changes of coordinates at first significant order,
while here higher orders effect are considered and are actually
relevant in obtaining a more radical simplification of the LdG
functional. In terms of the Poincar\'e theory, this would
correspond to a ``further normalization'' of Poincar\'e normal
forms (see also Appendix B in this respect). We avoid entering in
a discussion of the fine details of this procedure thanks to the
fact our computations are completely explicit, and our ``brute
force'' approach -- made possible by dealing with a concrete case
rather than with a general theory {(and the use of symbolic
manipulation programs)} -- allows to avoid mathematical subtleties
(see \cite{Gae02,RNF} for a discussion of these in terms of
dynamical systems).

\section{The SO(3) adjoint representation}
\label{sec:so3}

In this section we give some algebraic details on the SO(3)
adjoint representation and its invariants and covariants of low
order. These will be of use in our subsequent computations.

\subsection{Definition and generators}

Let us consider the action of $G=SO(3)$ on the space $\mathcal{M}$
of $3 \times 3$ symmetric traceless matrices \beq Q \ = \
\begin{pmatrix} x_1 & x_2 & x_3 \\ x_2 & x_4 & x_5 \\ x_3 & x_5 &
- (x_1+x_4) \end{pmatrix} \ ; \eeq we recall that if $Q$ is such a
matrix, the element $R \in G$ acts on it by $Q \to R Q R^{-1}$. As
for the infinitesimal $SO(3)$ action, this is generated by {\small
\beq
 L_1 = \begin{pmatrix} 0 & 0 & 0 \\ 0 & 0 & -1 \\ 0 & 1 & 0  \end{pmatrix} , \
 L_2  =  \begin{pmatrix} 0 & 0 & 1 \\ 0 & 0 & 0 \\ - 1 & 0 & 0 \end{pmatrix} , \
 L_3 = \begin{pmatrix} 0 & -1 & 0 \\ 1 & 0 & 0 \\ 0 & 0 & 0  \end{pmatrix}  .
\eeq
}
These satisfy of course the $so(3)$ Lie algebra relations
\beq [L_i , L_j ] \ = \ \epsilon_{ijk} \ L_k \ . \eeq

\subsection{Invariants}
\label{sec:invariants}

The orbits of this $G$-action are three-dimensional, and are
indexed by \beq\label{eq:invar} T_2 \ = \ \Tr (Q^2) \ , \ \ T_3 =
\Tr (Q^3 ) \ ; \eeq the trace of $Q$ is also an invariant, but in
this case a trivial one, as by definition $Q \in \mathcal{M}$ implies $\Tr
(Q) = 0$. \footnote{As for the determinant of $Q$ and its
iterates, these are of course also invariant but are expressed as
a polynomial in terms of the traces: e.g. for a
three-dimensional matrix $A$ one has  $\mathtt{Det} (A) =  (1/3)
\mathtt{Tr} (A^3) - (1/2) \mathtt{Tr} (A^2) \mathtt{Tr} (A) +
(1/6) [ \mathtt{Tr} (A)]^3$.}

We also recall that in this case all orbits pass through the set
of diagonal matrices, and of course traces are invariant under
conjugation. This helps in checking identities or inequalities
among invariants, e.g. the basic relation \cite{GLJ,AlL} \beq
\label{eq:isotr} (T_2)^3 \ - \ 6 \, (T_3)^2 \ = \ 2
\, (\la_1  - \la_2 )^2 \, (2 \la_1 + \la_2)^2 \, (\la_1 + 2
\la_2)^2 \ \ge \ 0 \ ,  \eeq where $\la_1 , \la_2$ are the
eigenvalues of $Q$. Note that for $\la_1 = \la_2$, i.e. in the
uniaxial case, we have an equality; and conversely the equality
holds only in the uniaxial case.

\begin{figure}
  \includegraphics[width=180pt]{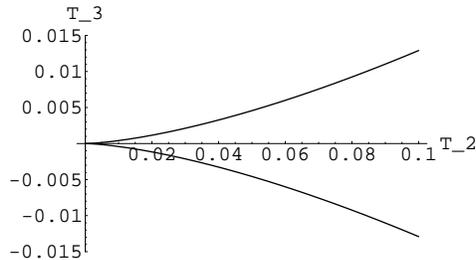}\\
  \caption{The allowed phase space in terms of invariants. The inequality \eqref{eq:isotr}
  implies the state of the system belongs to the region between the two curves
  $T_3 = \pm \sqrt{T_2^3 / 6}$. The borders of this region correspond to $\om = 0$, i.e.
  to the uniaxial phase, the interior to $0 < \om < 1$ to the biaxial phase.}\label{fig:fig1}
\end{figure}

It may be useful to have explicit expressions for the invariants
in terms of the $x_i$; with trivial computations these turn out to
be
\begin{eqnarray}
T_2 &=& x_1^2 \ + \ x_2^2 \ + \ x_3^2 \ + \ x_4^2 \ + \ x_5^2 \ + \ x_1 \, x_4 \ ; \label{eq:t2} \\
T_3 &=& x_1 \ (x_2^2 - x_4^2 - x_5^2) \ - \ x_4 \ (x_1^2 - x_2^2 + x_3^2 ) \ + \ 2 \, x_2 \, x_3 \, x_5  \ . \label{eq:t3} \end{eqnarray}

It should be noted that, by definition, the invariants and hence
the LdG functional are invariant under conjugation by SO(3)
matrices; thus one could aim at working with matrices in diagonal
form. In this context, a particularly convenient parametrization
is provided by $(q,\om)$, where $q \ge 0$ is the amplitude of the
tensorial order parameter \beql{eq:q} q \ = \ |Q| \ = \ \sqrt{T_2}
\eeq (see eq.\eqref{eq:Qnorm} below), and $\omega \in [0,1]$ is a
measure of the biaxiality \cite{GLJ,AlL}, defined in the present
notation by \beq\label{eq:om} \om \ = \ 1 \ - \ \sqrt{ \frac{6 \
T_3^2}{T_2^3} } \ . \eeq

Thus, the other three variables needed to complete a coordinate system beside
$(q,\om )$ could be seen as irrelevant ones (being coordinates
along the group orbits); note that the relation \eqref{eq:isotr} just
requires $q \ge 0$, $0 \le \om \le 1$.

For our approach it is however convenient to operate with the
$x_i$ variables, as degrees (orders) are defined in terms of these. Any
result given in terms of $T_2, T_3$ is promptly mapped to the
$(q,\om)$ variables formulation by recalling that
\beq\label{eq:qomega} T_2 \ = \ q^2 \ , \ \ T_3 \ = \ \frac{(1-\om
) \, q^3}{\sqrt{6}} \ . \eeq

{\medskip\noindent {\bf Remark 1.} Finally, we note that beside
the $SO(3)$ invariance mentioned above, the functions $T_2$ and
$T_3$ are also invariant under the discrete transformations
\begin{eqnarray*}
(x_1,x_2,x_3,x_4,x_5) & \to & (x_4, x_2 , \pm x_5 , x_1 , \pm x_3) \ , \\
(x_1,x_2,x_3,x_4,x_5) & \to & (x_1,-x_2,-x_3,x_4,x_5) \ , \\
(x_1,x_2,x_3,x_4,x_5) & \to & (x_1,-x_2,x_3,x_4,-x_5) \ . \end{eqnarray*}
Correspondingly, we have subspaces which are invariant under the gradient dynamics of any invariant potential (that is, the gradient at points of the subspace is granted to be tangent to the subspace), given by $\{ x_1 = x_4, x_3 = \pm x_5\}$, by $\{ x_2 = 0, x_3 = 0 \}$, and by $\{ x_2 = 0 , x_5 = 0\}$. The intersection $\{ x_2 = x_3 = x_5 = 0, x_4 = x_1 \}$ of these is also an invariant (one-dimensional) subspace.}

\subsection{Five-dimensional representation}

The adjoint $SO(3)$ action can be described by a linear (vector)
representation on the space $\^\mathcal{M} = \R^5 = \{ x_1,x_2,x_3,x_4,x_5
\}$; the matrices corresponding to the adjoint action of the
$SO(3)$ generators $L_i$ are {\small
$$
J_1 \, = \, \begin{pmatrix} 0 & 0 & 0 & 0 & 0 \\
 0 & 0 & -1 & 0 & 0 \\
 0 & 1 & 0 & 0 & 0 \\
 0 & 0 & 0 & 0 & -2 \\
 1 & 0 & 0 & 2 & 0 \end{pmatrix} \ , \
J_2 \, = \, \begin{pmatrix} 0 & 0 & 2 & 0 & 0 \\
 0 & 0 & 0 & 0 & 1 \\
 -2 & 0 & 0 & -1 & 0 \\
 0 & 0 & 0 & 0 & 0 \\
 0 & -1 & 0 & 0 & 0  \end{pmatrix} \ , \
J_3 \, = \, \begin{pmatrix}  0 & -2 & 0 & 0 & 0 \\
 1 & 0 & 0 & -1 & 0 \\
 0 & 0 & 0 & 0 & -1 \\
 0 & 2 & 0 & 0 & 0 \\
 0 & 0 & 1 & 0 & 0 \end{pmatrix} \ . $$
} It is immediate to check these are not orthogonal with respect
to the standard metric in $\R^5$. They are however orthogonal with
respect to the natural metric in the space of $n$-dimensional
matrices (here $n=5$), identified by $$ \langle A , B \rangle \ =
\ \frac{1}{n} \ \mathtt{Tr} (A^+ B ) \ ; $$ note that with this
metric, \beql{eq:Qnorm} |Q|^2 \ = \ \langle Q , Q \rangle \ = \
T_2 (x) \ = \ \sum_{i,j} Q_{ij}^2 \ . \eeq It would thus be
possible, at least in principles \footnote{This would actually
find some relevant obstacles when dealing with the physically
significant problem discussed in Sects. \ref{sec:unibireg} and
\ref{sec:unibising}; what we actually (implicitly) use there is a
more refined version of the normalization algorithm \cite{RNF},
and resorting to explicit allows to avoid discussing the general
mathematical theory, for which the reader is referred to
\cite{CGs,Gae02,RNF}.}, to use the general approach provided in
\cite{GL1}; we will however prefer to operate by direct explicit
computations.

\subsection{Covariants}

We are interested in explicitly determining the nonlinear
covariants for this action; that is, five-dimensional vectors
${\bf F}_k(x)$, homogeneous of degree $k$ in the $x_i$ -- hence
${\bf F}_k (a x) = a^k {\bf F}_k (x)$ -- which transform in the
same way as $\xb$, i.e. according to the same five-dimensional
representation of $SO(3)$. \footnote{In the mathematical
literature, these are also mentioned as 2-covariants, where the
``2'' specifies the SO(3) representation they follow; we recall
vector SO(3) representations have dimension $d = 2 \ell + 1$ with
$\ell \in {\bf Z_+}$, and in this case $\ell = 2$.}

This means satisfying at first order in $\eps$ the condition \beq
(I \ + \ \eps \, J_\a ) \ {\bf F}_k (\xb) \ = \ {\bf F}_{{ k}}
(\xb + \eps J_\a \xb) \eeq for $\a = 1,2,3$; {i.e. $(I + \eps L )
{\bf F}_k (\xb) = {\bf F}_k (\xb + \eps L \xb)$ for any $L \in
so(3)$}. This is of course equivalent to \beq g [{\bf F} (x)] \ =
\ {\bf F} (g x) \ \ \ \forall g \in G = SO(3) \ . \eeq

The multiplicity of these can be determined using Molien functions
\cite{JB,Sat,Mic,MZZ,CL,Cho}, see Appendix A. Here it is not
enough to know the number of covariants (and invariants), but we
need their explicit expressions; they can be determined in several
ways, including by direct explicit computations.

It turns out that there is a single covariant for each of the
orders $k=1$, $k=2$ and $k=3$, while there are two covariants for
each of the orders $k=4$, $k=5$ and $k=6$; there are three
covariants at each of the orders $k=7$ and $k=8$.
\footnote{Homogeneous covariants are determined up to linear
combinations; thus one could choose expressions with different
overall constants and, in the case where more covariants of the
same order exist, different linear combinations of them.}

Some of these covariants are rather obvious: e.g., the (only)
covariant of order one coincides with $\xb$, and there are
covariants of order $k$ given by the product of an invariant of
order $(k-1)$ with $\xb$. That is, we have \beq
\begin{array}{c} \Fb_1 \ = \ \xb \ , \ \ \Fb_3 \ = \ T_2 \, \xb \ , \
\ \Fb_4^{(1)} \ = \ T_3 \ \xb \ , \ \ \Fb_5^{(1)} \ = \ (T_2)^2 \
\xb \ , \\ \Fb_6^{(1)} \ = \ T_2 T_3 \, \xb \ , \ \ \Fb_7^{(1)} \
= \ T_2^3 \, \xb \ , \ \ \Fb_7^{(2)} \ = \ T_3^2 \, \xb \ , \ \
\Fb_8^{(1)} \ = \ T_2^2 \, T_3 \, \xb \ .
\end{array} \eeq

Then one can check by explicit computations that the second order
covariant (whose existence is guaranteed by Molien function
computations) is given by \beq \Fb_2 \ = \ \begin{pmatrix}
(x_1^2 + x_2^2 + x_3^2 ) \, - \, 2 \, (x_1 x_4 + x_4^2 + x_5^2) \\
  3 (x_1 x_2 + x_2 x_4 + x_3 x_5) \\ 3 (x_2 x_5 - x_3 x_4 ) \\
    (x_2^2 + x_4^2 + x_5^2) \, - \, 2 \, (x_1^2 + x_3^2 + x_1 x_4)  \\
    3 (x_2 x_3 - x_1 x_5) \end{pmatrix}  \eeq
This entails that we also have the covariants \beq
\begin{array}{c} \Fb_4^{(2)} \ = \ T_2 \ \Fb_2 \ , \ \ \Fb_5^{(2)} \ = \ T_3 \
\Fb_2 \ , \ \ \Fb_6^{(2)} \ = \ T_2^2 \ \Fb_2 \ , \\
\Fb_7^{(3)} \ = \ T_2 \, T_3 \, \Fb_2 ;  \ \ \Fb_8^{(2)} \ = \
T_2^3 \, \Fb_2 \ , \ \ \Fb_8^{(3)} \ = \ T_3^2 \, \Fb_2 \ .
\end{array} \eeq No other covariant exist at these orders, as
guaranteed by the Molien function approach \cite{Mic,MZZ,CL}, or
by explicit computation. \footnote{It should be noted that, albeit
we computed all the low order covariants, such a complete
knowledge is not always needed: in fact, covariants index the
(covariant) changes of variables, and knowing only some of them
means that the considered changes of variable would not be the
most general ones. E.g., in the following we will see that
disregarding one of the fourth order covariants, e.g.
$\Fb_4^{(2)}$, would not harm our procedure.}

\section{The Landau-deGennes functional and its reduction. Order six}
\label{sec:LdG}

It follows from our discussion in Section \ref{sec:so3}, see in
particular eqs.\eqref{eq:t2} and \eqref{eq:t3}, that the most
general invariant polynomial of order six can henceforth be
written {(with $c_i$ arbitrary constants)} as \beql{eq:landauT}
\Phi \ = \ c_1 \, T_2 \, + \, c_2 \, T_3 \, + \, c_3 \, T_2^2 \, +
\, c_4 \, T_2 \, T_3 \, + \, c_5 \, T_2^3 \, + \, c_6 \, T_3^2 \ .
\eeq In the following we will suppose to be {\it not } at exactly
the transition point, i.e. we will (until Sect.
\ref{sec:unibising}) assume \beql{eq:c1not0} c_1 \ \not= \ 0 \ .
\eeq {Note this means we are requiring the quadratic part of the
potential to be non-degenerate. As discussed and emphasized in
\cite{GL1}, this is an essential (and natural) condition for the
standard Poincar\'e approach to work \footnote{{In mathematically
more precise terms, we should actually require that $|c_1|$ is
large enough, with $|c_1| > \eps_*$ with $\eps_*$ depending on the
region of the phase space we want to study.}} .}

It is sometimes convenient to write $\Phi$ as \beql{eq:landauP1}
\Phi \ = \ \sum_{k=0}^4 \ \Phi_k \ , \eeq where $\Phi_k$ is
homogeneous of order $(k+2)$; needless to say for the $\Phi$ of
\eqref{eq:landauT} we have \beql{eq:landauP2}
\Phi_0 \ = \ c_1 \, T_2 , \ \Phi_1 \ = \ c_2 \, T_3 , \ \Phi_2 \ = \ c_3 \, T_2^2 , \ \
\Phi_3 \ = \ c_4 \, T_2 \, T_3 , \ \Phi_4 \ = \ c_5 \, T_2^3 \ + \ c_6 \, T_3^2 \ . 
\eeq

The central idea in the Poincar\'e approach \cite{ArnG,Elp,CGs} is
to perform near-identity changes of variables $x \to y = x + \eps
h (x)$. Note that here the small parameter $\eps$ does not need to
be explicit; actually, it should be seen as the size of the region
in which we operate. In other words, we want to consider changes
of variables of the form \beql{eq:cov} \xb \ \to \ \yb \ = \ \xb \
+ \ {\bf h} (\xb ) \ , \eeq where $\hb$ is at least quadratic in
the $x_i$.

As we want to deal with polynomials, $\hb$ should itself be a
polynomial; moreover in Landau theory symmetry is a central
ingredient of the theory, and hence we should make sure that the
change of variables preserves the symmetry.


All in all, this means that we should look for changes of
variables of the form \eqref{eq:cov} with $\hb$ a linear
combination of the higher order covariants identified above. That
is, we set
\beql{eq:covh} \hb \ = \ k_1 \, \Fb_2 \, + \, k_2 \, \Fb_3 \, + \,
k_3 \, \Fb_4^{(1)} \, + \, k_4 \, \Fb_4^{(2)} \, + \, k_5 \, \Fb_5^{(1)} \, + \,
k_6 \, \Fb_5^{(2)} 
\, .  \eeq

Note that, contrary to what is discussed in our general approach
described in \cite{GL1}, here we are {\it not} working step by
step, but just proceed by a brute force comprehensive computation;
this is possible in that we know {\it apriori} we just want to go
to order six (or order eight in sect. \ref{sec:red8}), and this
order is not too high.

\section{Poincar\'e transformations on the LdG functional of order six}
\label{sec:red6}

We will thus consider the general Landau polynomial $\Phi (\xb)$
and operate on it via the change of variables \eqref{eq:cov},
\eqref{eq:covh}. This will produce a new Landau polynomial
$\^\Phi$, which can again be written (with truncation at order six
in the $x_i$ variables) in the form \eqref{eq:landauP1}, i.e. as
\beq \^\Phi \ = \ \sum_{k=0}^4 \^\Phi_k \ . \eeq

\subsection{Generalities}

The expression of the $\^\Phi_k$ will depend both on the
expression of the original homogeneous polynomials $\Phi_m$, hence
on the coefficients $c_j$, and on the coefficients { $k_i$}
appearing in \eqref{eq:covh}. Note that the former have physical
significance and are given (at least for given values of the
external physical parameters) in our theory, but the latter are
just indexing the change of variables and we can choose them.

We will thus try to choose these in such a way to simplify as much
as possible the resulting Landau polynomial $\^\Phi$. In
particular, we will see that a suitable choice of the $k_i$ leads
to $\^\Phi_k (x) = 0$ for $k=1,...,4$; our simplification should
however preserve the stability of the theory, see next subsection
below.

We write $$ \Phi (\yb) \ = \ \Phi [\xb + \hb (\xb)] \ := \ \^\Phi
(\xb) $$ in terms of the explicit form of $\hb$ provided by
\eqref{eq:covh} and by the explicit expressions for the $\Fb_k$;
this produces explicit expressions for the $\^\Phi_k (\xb)$.

By construction, we will get $$ \^\Phi_0 (\xb ) \ = \ \Phi_0 (\xb)
\ , $$ as follows from the change of variables being a
near-identity one; higher order terms will be affected by our
change of variables. In any case, having chosen a covariant change
of coordinates \eqref{eq:cov} guarantees the $\^\Phi_k (\xb )$
will be invariant, hence written in the same form as in
\eqref{eq:landauP2} (obviously with different coefficients); that
is, we have necessarily \beql{eq:landauP3} \^\Phi_1 \ = \
\ga_3 \, T_3 , \ \  \^\Phi_2 \ = \ \ga_4 \, T_2^2 , \ \ \^\Phi_3 \
= \ \ga_5 \, T_2 \, T_3 , \ \  \^\Phi_4 \ = \ \ga_6 \, T_2^3 \ + \
\ga_7 \, T_3^2 \ .  \eeq The expressions for the coefficients
$\ga_k$ can be computed explicitly by simple albeit increasingly
involved algebra; it is convenient to perform these on a computer
using a symbolic manipulation language.

\subsection{Maximal order terms and convexity}
\label{sec:convex}

In dealing with the maximal order terms -- that is, the terms of
the order $N$ at which we truncate the Landau polynomial -- some
care should be taken. In fact, these terms control the
thermodynamical stability of the theory \cite{L1,Lan5}. The
criterion to ensure such a stability is that the Landau polynomial
should be convex for large (absolute) values of the order
parameter; in our case, this means large $|Q|$. In our case, and
more generally when the Poincar\'e transformations would be able
to completely cancel terms of this order and $N$ is even, a simple
way to guarantee the criterion is satisfied is by having a highest
order term of the type $\Phi_N = |Q|^{2q} = \rho^q$, where $q
=(N+2)$, and of course $\rho = |Q|^2$. See also \cite{GL1}.

\subsection{The simplifying transformation. I: non-maximal orders}

We will now implement the procedure described above. Thus at first
order we get \beql{eq:phi1} \^\Phi_1 \ = \ (c_2 \ + \ 9 \, c_1 \,
k_1 ) \ T_3 (x)  \ . \eeq Obviously, it suffices to choose
\beql{eq:k1} k_1 \ = \ - \, \frac{c_2}{9 \, c_1} \ , \eeq which is
possible thanks to \eqref{eq:c1not0}, in order to get $\^\Phi_1 =
0$, i.e. eliminate the cubic terms from the Landau potential.

The quartic term reads \beq \^\Phi_2 \ = \ (1/3) \, (c_3 + 2 c_2
k_1 + 3 c_1 k_1^2 + 2 c_1 k_2) \ [T_2 (x)]^2 \ ; \eeq by choosing
$k_1$ according to \eqref{eq:k1} this reduces to \beql{eq:phi2}
\^\Phi_2 \ = \  \( 3 c_3 - \frac{5 c_2^2}{27 c_1} + 2 c_1 k_2 \)
\,  [T_2 (x)]^2 \ . \eeq It thus suffices to choose \beql{eq:k2}
k_2 \ = \ \frac{5 \, c_2^2 \ - \ 27 \, c_1 \, c_3}{54 \, c_1^2}
\eeq to get $\^\Phi_2 (x) = 0$ as well. Note we are again using
\eqref{eq:c1not0}; this will also be true in the next steps, but
we will not remark it any more.

The term of order five turns out to be
\begin{eqnarray} \^\Phi_3 &=& (c_4 + 18 c_3 k_1 + 9 c_2 k_1^2 + 3 c_2 k_2 +
9 c_1 k_1 k_2 + 2 c_1 k_3 + 9 c_1 k_4)  \ T_2 (x) \ T_3 (x) \ ; \end{eqnarray}
choosing $k_1$ and $k_2$ as in \eqref{eq:k1} and \eqref{eq:k2}, this reduces to
\begin{eqnarray}  \^\Phi_3 &=& \frac{1}{27 \, c_1^2} \
\[ 8 c_2^3 - 81 c_1 c_2 c_3 + 27 c_1^2 (c_4 + c_1 (2 k_3 + 9 k_4)) \] \  T_2 (x) \ T_3 (x) \ . \end{eqnarray}
Here we have two parameters, $k_3$ and $k_4$, which are not yet
determined. By choosing e.g. \beql{eq:k3} k_3 \ = \ \frac{- 8
c_2^3 + 81 c_1 c_2 c_3 - 27 c_1^2 c_4 - 243 c_1^3 k_4}{54 \,
c_1^3} \  \eeq (note we are free to set $k_4 = 0$ if desired), or
however $k_3$ and $k_4$ satisfying
$$ 2 \, k_3 \ + \ 9 \, k_4 \ = \  \frac{- 8 c_2^3 + 81 c_1 c_2 c_3 - 27 c_1^2 c_4}{27 \, c_1^3} \ , $$
we get $\^\Phi_3 = 0$.

In this way, under the assumption $c_1 \not= 0$, we have reduced
the original Landau potential \eqref{eq:landauT} to the simpler
form \beql{eq:landauRed} \^\Phi (x) \ = \ c_1 \, T_2 (x) \ + \ \a
\, [T_2 (x)]^3 \ + \ \b \, [T_3 (x)]^2 \ ; \eeq here $c_1$ is the
same coefficient as in the original potential, while $\a$ and $\b$
are coefficients depending on the original coefficients $c_i$ as
well as on the coefficients $k_i$ entering in the function ${\bf
h} (\xb )$ identifying the transformation, see \eqref{eq:covh}.

By explicit computations, these are (recall $k_4$ is
undetermined and can be set to zero if desired)
\begin{eqnarray}
\a &=& [-29 c_2^4 + 558 c_1 c_2^2 c_3 - 1701 c_1^2 c_3^2 - 216 c_1^2 c_2 c_4  +
972 c_1^3 c_5 + 1296 c_1^3 c_2 k_4 ] \, / \, [972 \ c_1^3] \ ; \label{eq:alpha}  \\
\b &=& [ -c_2^4 + 12 c_1 c_2^2 c_3 - 6 c_1^2 c_2 c_4 + 3 c_1^3 c_6 - 27 c_1^3 c_2 k_4 ] \, / \,
[3 \ c_1^3] \ . \label{eq:beta} \end{eqnarray}

\subsection{The simplifying transformation. II: maximal order}

We have now to deal with terms of maximal order. Thus, we should
{\it not} try to set $\a = \b = 0$, and hence $\^{\Phi}_4=0$, but
rather try to simplify this term while being guaranteed it remains
convex for large $|Q|$.

It should be noted that the natural norm for the tensorial order
parameter $Q$ is just the one induced by the natural scalar
product in ${ \mathcal{M}} = GL(3)$, i.e. \beq |Q|^2 \ = \ \langle
Q , Q \rangle \ = \ \frac{1}{3} \ \mathtt{Tr} (Q^+ Q) \ ; \eeq
thus we have simply \beq |Q|^2 \ = \ \frac{1}{3} \ T_2 (x) \ .
\eeq

In other words, convexity is guaranteed if we have $\Phi_4 = a^2
[T_2 (x)]^3$ for any nonzero real number $a$; we will just set
$a=1$. This means looking for a transformation, i.e. for
coefficients $k_5$ and $k_6$ identifying the transformation, which
maps $(\a,\b)$ above into $\wt{\a} = 1$, $\wt{\b} = 0$.

This is obtained by choosing (the formulas can be slightly
simplified by a suitable choice of the value of the undetermined
parameter $k_4$)
\begin{eqnarray}
k_5 &=& \[ 29 c_2^4 - 558 c_1 c_2^2 c_3 + 27 c_1^2 (63 c_3^2 + 8 c_2 c_4) +
324 c_1^3 (3 - 3 c_5 - 4 c_2 k_4) \] \ \[ 1944 c_1^4 \]^{-1} \ , \label{eq:k5a} \\
k_6 &=& \[ c_2^4 - 12 c_1 c_2^2 c_3 - 3 c_1^3 c_6 + 3 c_1^2 c_2 (2 c_4 + 9 c_1 k_4) \] \
\[ 27 c_1^4 \]^{-1} \ . \label{eq:k6a} \end{eqnarray}

We have thus reduced the Landau polynomial of order six to the
form \beq \label{eq:landred6} \^{\Phi} (x) \ = \ c_1 \, T_2 (x) \
+ \ [T_2 (x)]^3 \ . \eeq The transformation producing this
simplification has been made completely explicit, being encoded in
the coefficients $k_1 , ... , k_6$.

Finally we note that, as discussed in \cite{GL1}, we should pay
some attention to the requirement the resulting series is a well
ordered one; as the denominators $\de_m$ appearing in the formula
for $k_m$ grow as $c_1^m$, we should require $c_1 > \eps$. In
other words, the radius of convergence of our transformation
(acting on the order parameters $\xb$) is estimated by $|c_1|$,
i.e. by the distance from the transition point $c_1 = 0$: the
transformation allow to deal with the Landau polynomial in a
simpler form provided we do not get too near to the transition
point.

Thus we can use it to analyze the possible phases at a given
nonzero value of the leading parameter $c_1$ (and possibly
secondary phase transitions), but not to analyze the situation
(e.g. compute the critical exponents) at the primary transition
point $c_1 = 0$. This problem will be tackled in
Sect.\ref{sec:unibising}.

\section{Simplification of the LdG functional at order eight}
\label{sec:red8}

We will now consider what happens if we have to deal with a higher
order, i.e. order eight, functional; one purpose of this section
is to show that even in this case one can obtain completely
explicit formulas. \footnote{Dealing with still higher orders does
not present any new conceptual difficulty, but the algebraic
manipulations become rapidly quite  involved.}

More relevantly, when in next section we will discuss the physical
implications of our computations, we will find that going to order
eight does remove an un-physical degeneration obtained at order
six, and gives results in agreement with those obtained (through
different methods) in the literature \cite{AlL}.

In this case we can still attempt a full cancellation of the terms
of order six and seven while the highest (eight) order terms
should not be fully eliminated but just simplified, for the same
reason as discussed above.

\subsection{Invariants, covariants, generating functions}

We have now to consider also higher order polynomials, so that
\eqref{eq:landauT} will be replaced by
\begin{eqnarray} \Phi &=& c_1 \, T_2 \, + \, c_2 \, T_3 \, + \, c_3 \, T_2^2 \, + \,
c_4 \, T_2 \, T_3 \, + \, c_5 \, T_2^3 \, + \, c_6 \, T_3^2 
 \, + \, c_7 \, T_2^2 \, T_3 \, + \, c_8 \, T_2 \, T_3^2 \, + \,
 c_9 \, T_2^4 . \label{eq:landauT2} \end{eqnarray}
Correspondingly, we will still consider a change of variables of
the form \eqref{eq:cov} but \eqref{eq:covh} will now be replaced
by
\begin{eqnarray} \hb &=& k_1 \, \Fb_2 \, + \, k_2 \, \Fb_3 \, + \,
k_3 \, \Fb_4^{(1)} \, + \, k_4 \, \Fb_4^{(2)} \, + \, k_5 \,
\Fb_5^{(1)} \, + \, k_6 \, \Fb_5^{(2)} \nonumber \\ & & \, + \,
k_7 \, \Fb_6^{(1)} \, + \, k_8 \, \Fb_6^{(2)}  \, + \, k_9 \,
\Fb_7^{(1)} \, + \, k_{10} \, \Fb_7^{(2)}\, + \, k_{11} \,
\Fb_7^{(3)} \ . \label{eq:covh2}\end{eqnarray}

\subsection{Terms of order six}

Needless to say, up to order five we get the same results as
above. The order six is not any more, in this framework, the
maximal one, hence we can attempt to fully eliminate it without
harming thermodynamical stability.

We know {\it apriori} that the invariant polynomial  will be of
the form \beq \^\Phi_4 (x) \ = \ \a \ [T_2 (x)]^3 \ + \ \b \ [T_3
(x)]^2 \ ; \eeq the values of the coefficients $\a$ and $\b$ --
provided $k_1$, $k_2$ and $k_3$ are chosen according to
\eqref{eq:k1}, \eqref{eq:k2} and \eqref{eq:k3} -- are given by
\eqref{eq:alpha}, \eqref{eq:beta}. It is possible to make both
$\a$ and $\b$ -- and hence $\^\Phi_4$ -- vanish by choosing (these
formulas can also be slightly simplified by a suitable choice of
the undetermined parameter $k_4$)
\begin{eqnarray}
k_5 &=& [29 c_2^4 - 558 c_1 c_2^2 c_3 + 243 c_1^2 (7 c_3^2 - 4 c_1 c_5)
 - 216 c_1^2 c_2 (-c_4 + 6 c_1 k_4)] \ [1944 \ c_1^4]^{-1} \ , \label{eq:k5} \\
k_6 &=& [ c_2^4 - 12 c_1 c_2^2 c_3 - 3 c_1^3 c_6 + 3 c_1^2 c_2 (2 c_4 + 9 c_1 k_4) ]
 \ [27 \ c_1^4]^{-1} \ . \label{eq:k6} \end{eqnarray}

\subsection{Terms of order seven}

At order seven we have only one invariant, i.e. we know that \beq
\^\Phi_5 \ = \ \ga \ T_2^2 \ T_3 \ . \eeq Setting $k_1,..., k_6$
as determined at lower orders, we get an explicit expression for
$\ga$ which depends moreover on the two parameters $k_7$ and
$k_8$; these can be chosen -- and actually one of these can be
chosen at will -- so to obtain $\ga=0$; details are given in
Appendix C. In other words, we can obtain $$ \^\Phi_5 \ = \ 0 \ .
$$

\subsection{Terms of order eight}
\label{sec:red8ord8}

At order eight, we are at maximal order and we should not fully
eliminate terms of this order; with the same discussion as in the
$N=4$ case of the previous section we see that they should instead
be reduced, if possible, to $\^{\Phi}_6 = [T_2 (x)]^4$.

Now we have two invariants of order eight, i.e. $T_2^4$ and $T_2
T_3^2$; thus we know {\it a priori} that, for whatever choice of
the $k_i$, we will have
$$ \^{\Phi}_6 (x) \ = \ \xi \, [T_2 (x)]^4 \ + \
\eta \, T_2 (x) \, [T_3 (x)]^2  $$ for some $\xi$, $\eta$; thus we
should aim at $$ \xi \ = \ 1 \ , \ \ \eta \ = \ 0 \ . $$

Explicit forms for $\xi$ and $\eta$ are easily computed (and
reported in Appendix C). The requirements $\xi = 1$, $\eta = 0$
can be satisfied by a suitable choice of the parameters $k_8$ and
$k_9$ (explicit for of these choices are given again in Appendix
C).

These present a new feature: while so far all the denominators
only depended on $c_1$, at this stage we will have denominators of
the parameters (and also of the resulting coefficients for
$\^\Phi_6$) which also depend on $c_2$.

Thus for the results to make sense we should require not only
$|c_1|$ large enough (which, as we remarked several time, is
inherent to the very spirit of the standard Poincar\'e approach),
but $|c_2|$ large enough as well.

On the other hand, if we only require $\xi = 1$ (without requiring
also $\eta = 0$), this can be satisfied with parameters -- and
resulting coefficients -- which do not see the appearance of $c_2$
in the denominator (see Appendix C).

With the choices described above, and with the same cautionary
notes about the need to have $|c_1|$ and $|c_2|$ large enough, we can
reduce the Landau polynomial to the form \beq \label{eq:landred8a}
\^\Phi (x)  \ = \ c_1 \, T_2 (x) \ + \ [T_2 (x)]^4 \ ; \eeq
needless to say what matters here are not the explicit (and rather
involved) expressions obtained for $k_1 , ... , k_9$, but the fact
such expressions {\it can } be explicitly determined and yield
\eqref{eq:landred8a}.

If we assume { $|c_1|$} large enough, but are not ready to make
any assumption regarding $c_2$, we can still reduce the Landau
polynomial to the form \beq \label{eq:landred8b} \^\Phi (x) \ = \
c_1 \, T_2 (x) \ + \ [T_2 (x)]^4 \ + \ \eta \ T_2 (x) \ [T_3
(x)]^2 \ . \eeq

\section{Biaxial and uniaxial nematic phases.
I: analysis of the simplified potential away from the main
transition point.} \label{sec:unibireg}

A physically relevant question is whether the theory allows for
direct transitions to biaxial phases, or if only transitions to
uniaxial ones are allowed directly from the fully isotropic phase
\cite{Fre}; see e.g. the discussion in \cite{AlL}. {This section
is devoted to applying our approach in this context. As already
mentioned our method cannot (without modifications, see next
section) deal with the degeneration corresponding to the phase
transitions. On the other hand, our method provides some hint,
consisting in the form of the would-be bifurcating branch after
the phase transition; this information will allow (see Sect.
\ref{sec:BUBS}) to identify an ansatz for the would-be biaxial
bifurcating solution and effectively run computations to provide
the first term in the series expansion of a biaxial branch
bifurcating directly from the fully symmetric solutions, and
determine its stability.}

It should be recalled that in the present notation anisotropy is
measured by the parameter $\om \in [0,1]$, defined in terms of
$T_2$ and $T_3$ by \eqref{eq:om}; see also Fig.\ref{fig:fig1}. The
biaxial phase has $\om >0$, while the uniaxial one is
characterized by $\om = 0$.

\medskip\noindent
{{\bf Remark 2.} A good deal of the discussion about this problem
present in the literature has been conducted using either the
$(q,\om)$ coordinates or the orbit space ones, i.e. $(T_2,T_3)$.
It should be stressed that from the point of view of perturbation
theory this causes troubles, in that one is destroying the grading
present in the $x$ coordinates; moreover, the branching point
$(q,\om) = (0,0)$ or $(T_2,T_3)=(0,0)$ is lying on the border of
the domain of definition. We will thus work in the $x$ coordinates
in order to reduce the LdG potential; once this is done, working
in the $(q,\om)$ or $(T_2,T_3)$ variables is legitimate.}

\subsection{{Sixth order potential}}

The result of the computations with a sixth order LdG potential is
that we can always reduce to a potential of the form
\eqref{eq:landred6}. When we look for critical points, these are
identified by \beq \nabla \^\Phi \ = \ \[ c_1 \ + \ 3 [T_2 (x)]^2
\] \ \nabla T_2 (x) \ = \ 0 \ . \eeq Thus all (and only, apart
from the trivial one $x=0$) the points satisfying \beql{eq:crit6}
T_2 (x) \ = \ |Q (x) |^2 \ = \ \sqrt{- \, c_1 / 3} \eeq are
critical ones. Needless to say, this is possible only for $c_1 <
0$; for $c_1 > 0$ the fully isotropic phase is stable.

The point is that \eqref{eq:crit6} does not depend at all on
$T_3$; thus the outcome of our computations at order six is that
all values of $T_3 (x)$ compatible with the value of $T_2 (x)$ in
view of \eqref{eq:isotr} would be allowed.

This makes little sense physically, and is in contrast with well
established results in the literature \cite{AlL}. As we are dealing
with a perturbation approach and the situation obtained at order six
is degenerate (no selection on $T_3$), it is natural to try to remove the
degeneration by going at higher orders.

\subsection{{Eight order potential}}
\label{sec:biaxial8}

Going at order eight the situation is indeed different. If we are
ready to make assumptions on $c_2$ -- beside those on $c_1$ --
then the situation is similar to the one described above, except
that $\^\Phi$ will be of the form \eqref{eq:landred8a} and hence
the critical points identified by \beq \nabla \^\Phi \ = \ \[ c_1
\ + \ 4 [T_2 (x)]^3 \] \ \nabla T_2 (x) \ = \ 0 \ . \eeq Thus all
(and only, apart from the trivial one $x=0$) the points satisfying
\beql{eq:crit8a} T_2 (x) \ = \ 3 \, |Q (x) |^2 \ = \ (- \, c_1
/4)^{1/3} \eeq are critical ones. { Note that now symmetry
breaking phases again are possible only for $c_1 < 0$, as
expected}.

On the other hand, if we want to deal with a generic $c_2$ we only
{ get} to $\^\Phi$ given by \eqref{eq:landred8b}; we assume $\eta
\not= 0$ (or we would be reduced to the previous case).

In the following we will also write $$ { c_1 \ = \ - \, \la} \ ,
$$ to emphasize this is a varying parameter (we will consider the
other ones as given) and that we are interested in the case $c_1 <
0$ (so that the origin is not a minimum). Thus
\eqref{eq:landred8b} reads now \beq \label{eq:landred8bb} \^\Phi
(x) \ = \ - \, \la \, T_2 (x) \ + \ \eta \ T_2 (x) \ [T_3 (x)]^2 \
+ \ [T_2 (x)]^4 \ . \eeq An explicit expression for $\eta$ is
computed in Appendix C.

\subsection{Analysis of the reduced potential}

With \eqref{eq:landred8bb}, critical points are identified by
 \beql{eq:gt8} \nabla \^\Phi \ =
\  \[ - \, \la \ + \ 4 \, [T_2 (x)]^3 \ + \ \eta \, [T_3 (x)]^2
\] \ \nabla T_2 (x) \ + \ \[ 2 \, \eta \, T_2 (x) \, { T_3 (x)} \]
\ \nabla T_3 (x) \ = \ 0 \ .  \eeq

Writing this explicitly in terms of the $x$ would produce a quite
involved equation, which cannot be easily handled; it is
convenient to analyze the problem in the form \eqref{eq:gt8}.

This equation requires the vanishing of a vector, which is
expressed as the sum of the two gradients $\nabla T_2$ and $\nabla
T_3$ with certain $x$-dependent coefficients. If the two gradients
are not collinear, the coefficients must vanish separately (the
gradients themselves are nowhere zero outside the origin), while
in case of collinearity the two vectors can combine to give a zero
sum. Thus in order to discuss solutions to \eqref{eq:gt8}, we
should distinguish two cases, i.e. points such that $\nabla T_3
(x) = \mu \nabla T_2 (x)$ for some real constant $\mu$, and points
such that $\nabla T_3 (x) \not= \mu \nabla T_2 (x)$ for any $\mu$.

Let us first consider the case where the two gradients are nowhere
collinear. In this case we must have \beql{eq:nocoll}
\begin{cases} c_1 \ + \ 4 \, T_2^3 \ + \ \eta \, T_3^2 \ = \ 0 & ,
\\ 2 \, \eta \ T_2 \, T_3 \ = \ 0 & ; \end{cases} \eeq recalling
that by hypothesis $\eta \not= 0$ \footnote{Note however that
$\eta = 0$ leaves us with $T_3$ fully undetermined (in fact, in
\eqref{eq:nocoll} $\eta$ and $T_3$ always appear together), and
$T_2^2 = q^2 = \la /4$.}, we have either $T_2 = 0$ or $T_3 = 0$.
The first case implies $|Q|^2 = 0$ and is thus not relevant.

In the second case, $T_3 = 0$, which implies $\om = 1$ and hence a
biaxial phase, the other equation yields \beql{eq:T20} T_2 = - (-
c_1 /4)^{1/3} \ = \ - \, (\la / 4)^{1/3} \ ; \eeq {for $\la > 0$
this entails $q^2 < 0$, and is thus not acceptable.}

We can now pass to consider the case where there are points ${\bf
x} \in \R^5$ such that the two gradients are collinear. The
explicit expressions for the gradients are easily obtained. With
lengthy explicit computations (performed with Mathematica and not
to be reported here), it turns out that when we require $ \nabla
T_3 = \mu \ \nabla T_2$ there are four possibilities for $\mu$,
i.e. (here and in the following, we write $\theta := \sqrt{ (x_1 -
x_4)^2 \, + \, 4 \, x_2^2 }$)
\beql{eq:mu}
\begin{array}{lll}
(a) & \mu = & - \, x_1 \ , \\
(b) & \mu = & - \, x_4 \ , \\
(c) & \mu = & - \, (1/2) \, [ x_1 + x_4 + \theta ] \ , \\
(d) & \mu = & - \, (1/2) \, [ x_1 + x_4 - \theta ] \ . \end{array} \eeq

Each of the $\mu$ values given in \eqref{eq:mu} allows for { two
(multi-dimensional)} branches of nontrivial critical points of the
LdG potential. We will now briefly discuss their features, and
identify one-parameter families of solutions embedded in such
branches. It should be stressed that such one-parameter families
fall within the scope of Remark 1; thus the restriction to these
is legitimate independently of the normalization procedure.

\begin{itemize}

\item[(a)] The case $(a)$ gives two two-dimensional branches,
\beql{eq:cpagen} x_2 = x_3  = 0 , \ x_5 = \pm \sqrt{2 x_1^2 - x_1
x_4 - x_4^2} \ . \eeq Note these are acceptable only for $- 2 x_1
\le x_4 \le x_1$. With \eqref{eq:cpagen}, we get immediately
\beql{eq:TA} T_2 \ = \ 3 \ x_1^2 \ , \ \ T_3 \ = \ - \, 2 \ x_1^3
\ ; \ \ \om \ = \ 1 \ - \ \frac{2 \ \sqrt{2}}{3} \ \approx \
0.057191 \ . \eeq

Simpler formulas are obtained by choosing special values for
$x_4$; e.g. we can choose $x_4 = 0$ and get the one-dimensional
representative branch $x_2 = x_3 = x_4 = 0$, $x_5 = \pm \sqrt{2}
x_1$; or choose $x_4 = x_1$, with also $x_2 = x_3= x_5 = 0$.

Inserting the conditions \eqref{eq:cpagen} into the equation
$\nabla \^\Phi = 0$ we get immediately that the latter yields
\beql{eq:xcrita} x_1 \ = \ \pm \ \( \frac{ \la }{4 \ (27 + 4
\eta)} \)^{1/6} \ , \eeq which is of course acceptable only for
either one of \beql{eq:xcritcond} \left\{ \la \ge 0 \ \mathrm{and}
\ \ \eta
> - 27/4 \ , \ \ \la \le 0 \ \mathrm{and} \ \ \eta < - 27/4
\right\} \ . \eeq

\item[(b)]
Similarly, the case $(b)$ also gives two two-dimensional branches,
 \beql{eq:cpbgen} \ x_2 = x_5 = 0 , \
x_3 = \pm \sqrt{ 2 x_4^2 - x_1 x_4 - x_1^2} \ . \eeq These are
acceptable only for $x_4 \le - x_1/2$ or $x_4 \ge x_1$.
With \eqref{eq:cpbgen}, we get immediately
\beql{eq:TB} T_2 \ = \ 3 \ x_4^2 \ , \ \ T_3 \ = \ - \, 2 \ x_4^3
\ ; \ \ \om \ = \ 1 \ - \ \frac{2 \ \sqrt{2}}{3} \ . \eeq

Simpler formulas are obtained choosing special values for $x_4$;
e.g. choosing $x_4 = x_1$ we get the one-dimensional branch
representative $x_2 = x_3 = x_5 = 0$, $x_4 = x_1$.

Inserting the conditions \eqref{eq:cpagen} into the equation
$\nabla \^\Phi = 0$ we get immediately that the latter yields
again \eqref{eq:xcrita}; the conditions \eqref{eq:xcritcond} do
still apply. Actually, the one-dimensional representatives
obtained for $(a)$ and $(b)$ are just the same.

\item[(c)]
As for $(c)$, in this case we get two three-dimensional branches,
\begin{eqnarray}  x_3 &=& \pm \frac{1}{\sqrt{2}} \,
\sqrt{2 x_2^2 + 2 x_4 (x_4 + \theta) - x_1 (x_1 + x_4 - \theta)} \ , \nonumber \\
x_5 &=& \pm \frac{- \theta - x_1 + x_4}{4 x_2} \, \sqrt{4 [x_2^2 +
x_4 (x_4 + \theta)] + 2 x_1 (\theta - x_4 - x_1) } \ .
\label{eq:cpcgen} \end{eqnarray} These  are acceptable provided
all the arguments of the roots are positive; we will not analyze
this condition in detail. The expressions for $T_2$ and $T_3$ are
readily obtained.

Formulas become simpler if we set $x_4 = x_1$, which implies
$\theta = 2 x_2$; in this case we get \beq T_2 = 3 (x_1 + x_2)^2 \
, \ T_3 = - 2 (x_1 + x_2)^3 \ ; \ \ \om = 1 \ - \ \frac{2 \
\sqrt{2}}{3} \ . \eeq The resulting two-dimensional branch is
identified by $x_3 = \pm \sqrt{x_2 (3 x_1 + x_2)}$, $x_5 = - x_3$;
for $x_1 >0$ this is acceptable for $x_2 \le - 3 x_1$ or $x_2 \ge
0$, while for $x_1< 0$ we require either $x_2 \le 0$ or $x_2 \ge -
3x_1$.

In both cases we can set $x_2 = x_1$, which gives a one
dimensional representative for this branch, $x_2 = x_1$, $x_3 =
\pm 2 x_1$, $x_4 = x_1$, $x_5 = \mp 2 x_1$. Inserting these into
the equation $\nabla \^\Phi = 0$, the latter yields again
\eqref{eq:xcrita}; the conditions \eqref{eq:xcritcond} do again
apply.

\item[(d)] Finally, the case $(d)$ gives two three-dimensional
branches as well (again we won't discuss in detail the bounds on
the admissible values),
\begin{eqnarray} x_3 &=& \pm \frac{1}{\sqrt{2}} \, \sqrt{2 x_2^2 + 2 x_4 (x_4 - \theta) - x_1 (x_1 + x_4 + \theta)} \ , \nonumber \\
x_5 &=& \pm \frac{\theta - x_1 + x_4}{4 x_2} \, \sqrt{4 [x_2^2 + x_4 (x_4 - \theta)] - 2 x_1 (\theta + x_4 + x_1) } \ . \label{eq:cpdgen} \end{eqnarray}

Here again we get simpler formulas by choosing $x_4 =  x_1$, which
yields \beq T_2 \ = \ 3 \ (x_1 - x_2)^2 \ , \ T_3 \ = \  - \, 2 \
(x_1 - x_2)^3 \ ; \ \ \om \ = \ 1 \ - \ \frac{2 \ \sqrt{2}}{3} \ .
\eeq The two-dimensional branch is identified by $x_3 = \pm
\sqrt{x_2 (x_2 - 3 x_1)}$, $x_5 = x_3$. For $x_1> 0$ this requires
either $x_2 < 0$ or $x_2 > 3 x_1$; for $x_1 < 0$ it requires
either $x_2 \ge 0$ or $x_2 \le 3 x_1$.

In both cases we can set $x_2 = - x_1$ and have a one-dimensional
representative of the branch, $x_2 = - x_1$, $x_3 = \pm 2 x_1$,
$x_4 = x_1$, $x_5 = \pm 2 x_1.$ Inserting these into the equation
$\nabla \^\Phi = 0$ the latter yields again \eqref{eq:xcrita}; the
conditions \eqref{eq:xcritcond} do again apply.
\end{itemize}
\bigskip

We have thus discussed the properties of these branches of
solutions, but not yet investigated their stability. In order to
do so, we can consider the Hessian $H_{ij} = (\pa^2 \^\Phi / \pa
x^i \pa x^j)$ along the solution branches. In this computation it
will be convenient to consider the one-dimensional representative
identified above.

Note that due to the dimensionality of the branches, we will
always have two  zero eigenvalues. Thus the condition of stability
is that the other three eigenvalues are positive.

\begin{itemize}

\item[(a-b)] In case $(a)$ we easily obtain the Hessian, which is
rather simple; its eigenvalues are \beq \sigma_{(a)} \ = \ \{ 0 \
, \ 0 \ , \ 18 \, \la \ , \ - \frac{18 \, \eta \, \la}{27 + 4
\eta} \ , \  - \frac{18 \, \eta \, \la}{27 + 4 \eta} \} \ . \eeq
Recalling \eqref{eq:xcritcond}, the denominator is always positive
for $\la \ge 0$; thus the condition for (existence and) local
stability of this branch is \beq \la \ > \ 0 \ , \ \ - 27/4 < \eta
< 0 \ . \eeq With our choice of the one-dimensional
representative, case (b) is identical to (a); hence we get the
same formulas and results.

\item[(c-d)] In the case $(c)$ we get a quite more complex
Hessian; its eigenvalues can still be computed and we get \beq
\sigma_{(b)} \ = \ \{ 0 \ , \ 0 \ , \ - \frac{12 \, \eta \,
\la}{27 + 4 \eta} \ , \ \frac{3 \, (486 + 47 \eta - \ga) \, \la}{8
\, (27 + 4 \eta)} \ , \ \frac{3 \, (486 + 47 \eta + \ga) \, \la}{8
\, (27 + 4 \eta)} \} \ , \eeq where we have written
$$ \ga \ = \ \sqrt{ 236196 \ + \ 94068 \, \eta \ + \ 9377 \eta^2 } \ ; $$
the argument of the square root is always positive. The last two
eigenvalues do always have the same sign as $\la$; thus again for
$\la>0$ this is a stable branch for $- 27/4 < \eta < 0$.
Computations and results are just the same in the case $(d)$ as
for the case $(c)$.

\end{itemize}

\bigskip\noindent
The present discussion suffices to conclude that in cases $(a) -
(d)$ we have branches of critical points with non-zero $\om$, i.e.
biaxial ones, and that they are stable \footnote{We stress we have
only studied \emph{local}  stability, determined by the Hessian;
our analysis does not exclude the possibility that $\Phi$ has also
different local minima, maybe with a lower value.} for $\la > 0$
provided $- 27/3 < \eta < 0$. Note that these are quite weakly
biaxial, as shown by the value of $\om = [1 - 2 \sqrt{2}/3]
\approx 0.057191$ (such a low value can pose serious problem for
experimentally distinguishing this biaxial solution from a
uniaxial one).

It should be stressed that our discussion does \emph{not} give a
proof of the possibility of direct transition from the fully
isotropic phase to biaxial phases (actually, in the next section
we will see things are quite different). In fact, our reduction
procedure fails precisely at the $c_1 = 0$ point.

The present analysis identifies branches of biaxial solutions
existing near -- but not ``too near'' -- to the critical point. If
there is a branch of biaxial solutions stemming directly from the
critical point, it must be of the same form, and thus our analysis
gives a hint for the form of the critical branch to be sought for
in this subsequent analysis, developed in the next section
\footnote{It should be stressed again that the $x_i$ appearing
above are those obtained {\it after} the normalization steps have
been carried out; this has to be taken into account in trying to
compute this solution branch in explicit terms. However, our
choice of one-dimensional representative has been guided by Remark
1, so that these are embedded in invariant subspaces for the full
Landau theory.}.

\section{Biaxial and uniaxial nematic phases. II: The main transition region}
\label{sec:unibising}

In the previous Section \ref{sec:unibireg} we have implemented our
method under the assumption $- c_1 = \la \not=0$. If we want to
consider the parameter region near the main phase transition at
$\la = 0$, that discussion does simply not apply.

In this section we will apply again our method keeping in mind we
want to analyze exactly the region near $c_1 = 0$. This will
produce some different results than for $c_1$ bounded away from
zero, and these results \emph{can } be used to analyze the biaxial
phase problem.

On the other hand, it is known that the analysis of the biaxial
phase problem leads to consider an intricate situation as
parameters are varied; in our case we will have less parameters
(after the simplification), but as the Physics has not changed we
should expect an equally intricate situation. The analysis of this
lies outside the limits of the present paper, so we will be
satisfied with showing that our method allows to identify a
simpler LdG potential via a change of variables which is
admissible in a full neighborhood (whose size will depend on
$|c_2| \not= 0$, see below) of the main transition point

In simplifying the LdG potential, we will work under the
assumption that the next-to-leading order term is nonzero, i.e.
\beql{eq:c2not0} c_2 \ \not= \ 0 \ ; \eeq this condition takes the
place of \eqref{eq:c1not0}.

\subsection{Simplified potential -- order six}

In order to analyze the phase transition taking place at $\la =
0$, we can implement the simplification procedure paying attention
to the fact that \emph{no division by a factor $c_1$ should take
place}.

We will first conduct our discussion based on the order six LdG
potential. We will thus write $\Phi$ and $\hb$ as in Sect.
\ref{sec:red6}, see \eqref{eq:landauT} and \eqref{eq:covh}
(actually we only need generators up to order four, i.e. in
\eqref{eq:covh} we can set $k_5=k_6=0$); that is, we have
(repeating here these formula for convenience of the reader)
$$ \Phi \ = \ c_1 T_2 + c_2 T_3 + c_3 T_2^2 + c_4 T_2 T_3 + c_5 T_2^3 + c_6 T_3^2 \
; \ \  \hb \ = \ k_1 \Fb_2 + k_2 \Fb_3 + k_3 \Fb_{41} + k_4
\Fb_{42} \ . $$

With these, we obtain a transformed potential $\^\Phi$ written in
the same way but with $\ga_i$ taking the place of $c_i$ (for $i
\not= 1$). By explicit computations we easily get the detailed
form of the $\ga_i$. These become simpler if we set, as we do in
the following, $k_1 = 0$; the reader will easily observe that no
further reduction would be possible by considering a nonzero $k_1$
(as we cannot divide by factors containing $c_1$). With this, we
get
\begin{eqnarray*}
\ga_2 &=& c_2  \ , \\
\ga_3 &=& c_3 \ + \ 2 c_1 k_2 \ , \\
\ga_4 &=& c_4 \ + \ 3 c_2 k_2 + 2 c_1 k_3 + 9 c_1 k_4 \ , \\
\ga_5 &=& c_5 \ + \ c_1 k_2^2 + 4 c_3 k_2 + 2 c_2 k_4 \ , \\
\ga_6 &=& c_6 \ + \ 3 c_2 k_3  \ . \end{eqnarray*}

It is now easy to choose the $k_i$ so to get a simpler form for
the LdG potential; in particular, we want to set either one of
\beql{eq:RPT6abc} \begin{cases} \ga_5 = 0 \ , \ \ \ga_6 = 1 & (a) \ , \\
\ga_5 = 1 \ , \ \ \ga_6 = 0 & (b) \ , \\
\ga_5 = 1 \ , \ \ \ga_6 = 1 & (c) \ ; \end{cases} \eeq actually in
all cases one can also set $\ga_4 = 0$, see below.

The reader can check that in all cases explicit solutions for the
critical points of the simplified potential $\^\Phi$ can be
explicitly obtained.

Case $(a)$ is obtained by choosing
\begin{eqnarray*}
k_2 &=& \frac{18 c_{2}^2-108
   c_{1}
   c_{3}-\sqrt{\left(108
   c_{1} c_{3}-18
   c_{2}^2\right)^2-108
   c_{1}^2 (27 c_{5}
   c_{1}+4 c_{6}
   c_{1}-4 c_{1}-6
   c_{2} c_{4})}}{54
   c_{1}^2} \\
   &=& -\frac{c_{4}}{3
   c_{2}}+\left(-\frac{2
   c_{3}
   c_{4}}{c_{2}^3}+\frac{3 c_{5}}{2
   c_{2}^2}+\frac{2
   c_{6}}{9
   c_{2}^2}-\frac{2}{9
   c_{2}^2}\right)
   c_{1}+O\left(c_{1}^
   2\right) \ ; \\
k_3 &=& \frac{1-c_{6}}{3
   c_{2}} \ ; \\
k_4 &=& \frac{-c_{1} k_{2}^2-4 c_{3} k_{2}- c_{5}}{2 c_{2}} \\
    &=& \left(\frac{2 c_{3} c_{4}}{3 c_{2}^2}-\frac{c_{5}}{2 c_{2}}\right)+
    \left(\frac{4 c_{4} c_{3}^2}{c_{2}^4}-\frac{3 c_{5}
   c_{3}}{c_{2}^3}-\frac{4 c_{6} c_{3}}{9 c_{2}^3}+\frac{4 c_{3}}{9 c_{2}^3}-
   \frac{c_{4}^2}{18 c_{2}^3}\right)
   c_{1}+O\left(c_{1}^2\right) \ .
\end{eqnarray*} With these, we obtain \beql{eq:RP6A} \^\Phi \ = \
- \la \, T_2 \ + \ \ga_2 \, T_3 \ + \ T_3^2 \ . \eeq The remaining
coefficients are given in terms of the original ones by
\beql{eq:RP6ga} \ga_2 \ = \ c_2 \ , \ \ \ga_3 \ = \ c_3 + 2 c_1
k_2 \ = \ c_3 \ + \ \frac{2 c_4}{3 c_2} \la \ + \ O (\la^2 ) \ .
\eeq

We obtain case $(b)$ by choosing
\begin{eqnarray*}
k_2 &=& \frac{18 c_{2}^2-108
   c_{1}
   c_{3}-\sqrt{\left(108
   c_{1} c_{3}-18
   c_{2}^2\right)^2-108
   c_{1}^2 (27 c_{5}
   c_{1}+4 c_{6}
   c_{1}-27 c_{1}-6
   c_{2} c_{4})}}{54
   c_{1}^2} \\
   &=& -\frac{c_{4}}{3
   c_{2}}+\left(-\frac{2
   c_{3}
   c_{4}}{c_{2}^3}+\frac{3 c_{5}}{2
   c_{2}^2}+\frac{2
   c_{6}}{9
   c_{2}^2}-\frac{3}{2
   c_{2}^2}\right)
   c_{1}+O\left(c_{1}^
   2\right) \ ; \\
k_3 &=& -\frac{c_{6}}{3 c_{2}} \ ; \\
k_4 &=& \frac{1-c_{1} k_{2}^2-4
   c_{3}
   k_{2}-c_{5}}{2
   c_{2}} \ . \end{eqnarray*}
With these, we obtain \beql{eq:RP6B} \^\Phi \ = \ - \la \, T_2 \ +
\ \ga_2 \, T_3 \ + \ \ga_3 \, T_2^2 \ + \ T_2^3 \ . \eeq The
remaining coefficients are again given in terms of the original
ones by \eqref{eq:RP6ga}.

Finally we note that case $(c)$ is obtained for
\begin{eqnarray*}
k_2 &=& \frac{18 c_{2}^2-108
   c_{1}
   c_{3}-\sqrt{\left(108
   c_{1} c_{3}-18
   c_{2}^2\right)^2-108
   c_{1}^2 (27 c_{5}
   c_{1}+4 c_{6}
   c_{1}-31 c_{1}-6
   c_{2} c_{4})}}{54
   c_{1}^2} \\
   &=& -\frac{c_{4}}{3
   c_{2}}+\left(-\frac{2
   c_{3}
   c_{4}}{c_{2}^3}+\frac{3 c_{5}}{2
   c_{2}^2}+\frac{2
   c_{6}}{9
   c_{2}^2}-\frac{31}{18
   c_{2}^2}\right)
   c_{1}+O\left(c_{1}^
   2\right) \ ; \\
k_3 &=& \frac{1-c_{6}}{3
   c_{2}} \ ; \\
k_4 &=& \frac{-3 c_{2}^4+18
   c_{1} c_{3}
   c_{2}^2+\sqrt{3}
   \sqrt{3 c_{2}^4-36
   c_{1} c_{3}
   c_{2}^2+6 c_{1}^2
   \left(18
   c_{3}^2+c_{2}
   c_{4}\right)+c_{1}^
   3 (-27 c_{5}-4
   c_{6}+31)}
   c_{2}^2-3 c_{1}^2
   c_{4} c_{2}+2
   c_{1}^3
   (c_{6}-1)}{27
   c_{1}^3 c_{2}} \\
   &=& \left(\frac{2 c_{3}
   c_{4}}{3
   c_{2}^2}-\frac{c_{5
   }}{2 c_{2}}+\frac{1}{2
   c_{2}}\right)+\left(\frac{4 c_{4}
   c_{3}^2}{c_{2}^4}-\frac{3 c_{5}
   c_{3}}{c_{2}^3}-\frac{4 c_{6} c_{3}}{9
   c_{2}^3}+\frac{31
   c_{3}}{9
   c_{2}^3}-\frac{c_{4}^2}{18 c_{2}^3}\right)
   c_{1}+O\left(c_{1}^2\right) \ . \end{eqnarray*}
In this case the potential reads \beql{eq:RP6both} \^\Phi \ = \ -
\la \, T_2 \ + \ \ga_2 \, T_3 \ + \ \ga_3 \, T_2^2 \ + \ T_3^2 \ +
\ T_2^3 \ , \eeq with the remaining coefficients given explicitly
by \eqref{eq:RP6ga} once again.

It is maybe worth remarking that having kept both terms of order
six but having fixed their coefficients to one, we are guaranteed
of convexity at large $|x|$; in fact, in terms of $(q,\om)$
variables we have
$$ T_2^3 \ + \ T_3^2 \ = \ \[ 1 \ + \ \frac{(1 - \om)^2}{6} \] \
q^6 \ . $$

The reader can note that $\ga_2$ and $\ga_3$, and actually also
$k_2$, are the same in the three cases; both facts are natural, in
that the three cases differ only for order six terms.

We have thus shown that, with explicit computations, the
Landau-deGennes potential can be reduced, \emph{uniformly in a
neighborhood of the transition point $c_1 = - \la = 0$} (again,
the size of this neighborhood being controlled by $c_2 \not= 0$)
to a simpler form, i.e. to either \eqref{eq:RP6A} or
\eqref{eq:RP6B} or \eqref{eq:RP6both}. Retaining the $\ga_2$ and
$\ga_3$ terms was unavoidable due to the requirement to avoid any
division by a $c_1$ factor.

\subsection{Simplified potential -- order eight}
\label{sec:SP8}

A similar analysis can be performed for the LdG potential of
degree eight, as suggested by the discussion of Sect.
\ref{sec:unibireg}, see \eqref{eq:landauT2}: in this case we
obtain more involved formulas, and several options are possible
concerning the simplification of the highest order term (i.e.
retaining the $T_2^4$ or the $T_2 T_3^2$ one; equivalently,
setting $\ga_8=0$, $\ga_9=1$ or $\ga_8=1$, $\ga_9=0$) and some of
the sub-maximal ones. It should be stressed that we can again
arrive at a reduced eight order potential for which explicit
(albeit rather involved) expressions for the critical points can
be obtained.

Here we will consider two possible forms of the reduced potential,
which correspond to the maximal possible simplification (in the
sense of eliminating as many terms as possible), i.e.
\beql{eq:RP8AB} \begin{cases} \^\Phi \ = \ - \la \, T_2 \ + \
\ga_2 \, T_3 \ + \ \ga_3 T_2^2 \ + \ \ga_5 \, T_2^3 \ + \ T_2 \,
T_3^2 & (a) \ , \\
\^\Phi \ = \ - \la \, T_2 \ + \ \ga_2 \, T_3 \ + \ \ga_3 T_2^2 \ +
\ \ga_6 \, T_3^2 \ + \ T_2^4 & (b) \ . \end{cases} \eeq

In case $(a)$ the remaining coefficients are given in terms of the
original ones by
\begin{eqnarray}
\ga_2 &=& c_2 \ , \nonumber \\
\ga_3 &=& c_3 \ + \ \frac{2
   c_{4}}{3 c_{2}} \, \la \ + \ \frac{4 c_{6}}{9
   c_{2}^2}\, \la^2 \ , \\
\ga_5 &=& \left(c_{5}-\frac{4 c_{3} c_{4}}{3
   c_{2}}\right) \ - \ \left(\frac{c_{4}^2}{c_{2}
   ^2}+\frac{16 c_{3} c_{6}}{9
   c_{2}^2}-\frac{2 c_{7}}{3 c_{2}}\right) \,
   \la \ + \ O\left( \la^2 \right) \ . \nonumber \end{eqnarray}

In case $(b)$ the remaining coefficients are given in terms of the
original ones by
\begin{eqnarray}
\ga_2 &=& c_2 \ , \nonumber \\
\ga_3 &=& c_3 \ + \ \frac{2 c_{4}}{3 c_{2}} \, \la \ , \\
\ga_6 &=& \left(c_6 \, - \, \frac{9 c_{3} c_{4}}{c_{2}} \, + \,
\frac{27
   c_{5}}{4} \right) \ - \ \left(\frac{27
   c_{4}^2}{4 c_{2}^2} \, - \, \frac{9 c_{7}}{2
   c_{2}}\right) \, \la \ + \ O\left(\la^2 \right) \ . \nonumber \end{eqnarray}

The coefficients of the change of coordinates taking us to these
expressions are rather involved and are given in Appendix D.

\subsection{Branching unstable biaxial solutions}
\label{sec:BUBS}

It would be natural to attempt to use these results to analyze the
biaxial phase problem; this would led us too far, but we present
here some computations based on the simplified potential and
related to the one-dimensional weakly biaxial branches determined
(outside the transition region) in Sect.\ref{sec:unibireg}. We
will show here that these branches are unstable at the transition
point.

It will suffice to consider the LdG potential at order six, i.e.
\eqref{eq:landauT}, with $\hb$ of the form \eqref{eq:covh}; we
will consider the reduced potential of the form \eqref{eq:RP6B},
i.e. \beql{eq:redpotbif} \^\Phi \ = \ - \, \la \, T_2 \ + \ \ga_2
\, T_3 \ + \ \ga_3 \, T_2^2 \ + \ T_2^3 \ , \eeq with actually $$
\ga_2 \ = \ c_2 ; \ \ \lim_{\la \to 0} \ga_3 \ = \ c_3 \ . $$

The discussion of Sect.\ref{sec:unibireg} suggests to focus on the
one-dimensional manifold $$ M \ = \ \{ x_1 = x , \ x_2 =  0 , \
x_3 = 0 , \ x_4 = x , \ x_5 = 0 \} \ . $$

Doing this, we are reduced to study a one-dimensional problem,
described by the effective potential \beq \Psi \ = \ - \, 3 \, \la
\ x^2 \ - \ 2 \, \ga_2 \ x^3 \ + \ 9 \, \ga_3 \ x^4 \ + \ x^6 \ .
\eeq We will write $\la = \eps$, and look for a solution as a
power series in $\eps$, i.e. as
$$ x \ = \ \sum_k z_k \ \eps^k \ . $$

With standard computations, we obtain at first orders
\begin{eqnarray*}
z_1 &=& - \, \frac{1}{\ga_2} \ ; \ \ z_2 \ = \ 6 \
\frac{\ga_3}{\ga_2^3} \ ; \ \
z_3 \ = \ - \, 72 \ \frac{\ga_3^2}{\ga_2^5} \ ; 
\ \ z_4 \ = \ 4 \ \( \frac{\ga_2^2 + 270 \ga_3^3}{\ga_2^7} \) \ ;
\ \ z_5 \ = \ - \, 144 \ \ga_3 \ \( \frac{\ga_2^2 +126
\ga_3^3}{\ga_2^9} \) \ .
\end{eqnarray*}

In order to study the stability of this solutions, we consider the
Hessian $H_b$ computed on the solutions branch so determined. This
will also be written as a series expansion, and we get
$$ H_b \ = \ H_1 \ \eps \ + \ H_2 \ \eps^2 \ + \ O (\eps^3) \ ; \ \ \ H_1 \ = \ \left(
\begin{array}{lllll}
 0 & 0 & 0 & 3 & 0 \\
 0 & -6 & 0 & 0 & 0 \\
 0 & 0 & 0 & 0 & 0 \\
 3 & 0 & 0 & 0 & 0 \\
 0 & 0 & 0 & 0 & 0
\end{array}
\right) \ ; \ \ H_2 \ = \ \frac{18 \ga_{3}}{\ga_{2}^2} \ \left(
\begin{array}{lllll}
 1 & 0 & 0 & 0 & 0 \\
 0 & 2 & 0 & 0 & 0 \\
 0 & 0 & 0 & 0 & 0 \\
 0 & 0 & 0 & 1 & 0 \\
 0 & 0 & 0 & 0 & 0
\end{array}
\right) \ . $$ As for the eigenvalues, we have a double zero
eigenvalue (which corresponds to the degeneracy of the problem, as
discussed in the previous section), and three nonzero ones, which
are \beq \mu_1 \ = \ - 6 \( 1 -  6 \frac{\ga_3}{\ga_2^2} \eps \)
\, \eps \ ; \ \ \mu_2 \ = \ - 3 \( 1  - 6 \frac{\ga_3}{\ga_2^2}
\eps \) \, \eps \ ; \ \ \mu_3 \ = \ 3 \( 1 + 6
\frac{\ga_3}{\ga_2^2} \eps \) \, \eps \ . \eeq

Thus for small $\eps$ (that is, small $\la$) we have an unstable
branch. For $\ga_3 > 0$, this becomes stable for \beq \la \ > \
\la_s \ \simeq \ \frac{\ga_2^2}{6 \ga_3} \ . \eeq

Some numerical experiments, conducted assigning random values to
$\ga_2$ and $\ga_3$, confirm this description. Moreover, they show
we always have a stable solution on the manifold $M$, but this is
\emph{not} branching off from the origin.

\section{Discussion and conclusions}
\label{sec:conc}

The Landau-deGennes potential describing the isotropic-nematic
phase transition (and those between different nematic phases) in
liquid crystals is a sixth order polynomial, depending on six
parameters, see \eqref{eq:landauT}; going to the next order, we
have an eight order parameter, depending on nine parameters, see
\eqref{eq:landauT2}. We have shown that passing to suitable
non-homogeneous new variables, see \eqref{eq:cov} and
\eqref{eq:covh2}, the potential is written in a simpler form, see
\eqref{eq:landred8b}, depending only on two parameters. The
transformation to reach this simpler form has been explicitly
determined in terms of the coefficients $c_i$ appearing in the
original potential.

This allows to study the equilibrium points of the potential in
the new variables, i.e. using the simplified form $\^\Phi (x)$.
Albeit going back to the original variables requires to invert the
nonlinear change of variables \eqref{eq:cov}, \eqref{eq:covh2} and
is thus a nontrivial task (but see below), qualitative information
obtained from the simplified potential remains true in whatever
coordinates.

The quantitative aspect would concern the value of the parameters
at which phase transitions occurs, and the relation between the
value of the parameters $c_i$ and those assumed by the order
parameters $x_i$. As mentioned above, in order to pass from the
new variables $x_i$ to the old ones one should invert the
nonlinear change of variables \eqref{eq:cov}, \eqref{eq:covh2};
note that it suffices to work at a finite order in $\eps = |\xb
|$, so that the inversion is obtained by a series expansion.
Explicit formulas could be obtained for the case at hand, but they
are very complex and not significant. \footnote{The conceptual
problems related to inversion is radically solved by considering
more refined changes of variables, corresponding to {\it Lie
series} \cite{CGs,Bro,BGG,Gio}. In this way, the map $y \mapsto x$
corresponds to the time-one flow of a vector field $X = h^i (x)
(\pa / \pa x^i)$, and inversion corresponds to time-reversed flow.
The computational details to actually obtain the inversion at
finite orders are however essentially the same. See \cite{CiW,Wency} for
other detail on convergence issues.}

{A significant weakness of our standard method, built in the basic
idea behind the Poincar\'e normalization approach, is that it
cannot be applied when the quadratic part of the Landau potential
vanishes. Or, this is precisely the situation met at the main
transition point, and one would be specially interested in
analyzing the transition region. Our method is however flexible
enough so that by a small modification -- consisting in avoiding
to operate division by factor which vanish at the transition point
-- it can be also applied around critical points (this makes that
low order terms cannot be eliminated, but several among the higher
order terms are anyway erased), as shown explicitly in
Sect.\ref{sec:unibising}).

We have moreover considered a concrete open problem, i.e. that of
possible direct transition from the fully isotropic phase to the
biaxial one \cite{Fre,AlL}. The implementation of the method in
the singular region as in Sect.\ref{sec:unibising} can be used to
analyze in simpler terms the possibility of having a stable
biaxial phase branching directly off the fully symmetric state,
and we hope to be able to tackle this problem in the near future.
Some partial result, concerning the special one-dimensional
branches identified by Remark 1 and studied in
Sect.\ref{sec:unibireg}, are presented in Sect.\ref{sec:BUBS}.}

In conclusion, we have shown by explicit computations that the
Poincar\'e approach, devised to study critical points of dynamical
systems, can also be profitably adopted to simplify computations
in Landau theory of phase transitions. We have here considered a
concrete and relevant application, i.e. the Landau-deGennes theory
for the isotropic-nematic transition in liquid crystals, but it is
clear that the validity of the Poincar\'e approach is much more
general.

It should also be mentioned that the LdG theory considered here
did {\it not} satisfy the simple hypotheses considered in
\cite{GL1} (under which quite general results were obtained) for
the standard metric in $\R^5$, but did for standard metric in
$GL(3)$. Rather than discussing things with this metric, here we
have implemented the essential of Poincar\'e ideas, i.e. looked
for a near-identity change of variables which does preserve the
symmetry of the problem at hand and depends on arbitrary
parameters; the latter can be chosen to obtain a simpler form of
the function under study (in this case the LdG potential) in the
new variables.

This also shows that the method proposed in \cite{GL1} can be
applied avoiding the (mild) mathematical sophistication which
would be needed to take into account the non-standard metric to be
introduced in the orbit space, see Appendix B.
The direct approach pursued here has also another advantage, also
discussed in Appendix B, i.e. that in this way
we are able to take into account the higher order effects which
were not considered in previous work \cite{GL1}.

\section*{Appendix A. The Molien function}

The Molien function \footnote{So called after the Latvian-born
Russian mathematician Theodor Molien (Riga 1861 -- Tomsk 1941);
the relevant work for our topic dates back to 1897.} is a
generating function; given a (``source'') representation $T$ of a
group $G$, acting on a vector space with variables $x_i$, the
coefficients of its expansion in series of the parameter yields
the number of independent tensors (with components being
polynomials in the $x_i$) which transforms under a possibly
different (``target'') representation $\wt{T}$ of $G$.

In the case of interest here, $G = SO(3)$, vector representations
are indexed by an integer number $\ell$ and have odd dimension $(2
\ell + 1)$ [spinor representations are indexed by half-integer
numbers $m/2$, with $m$ odd, and have even dimension $(m +1)$];
thus the trivial representation $\ell=0$ has dimension 1, the
defining representation $\ell=1$ has dimension 3, and the $\ell=2$
representation we are considering here has dimension 5. The Molien
function will then be denoted as $F_{\ell,\la} (t)$, where $t$ is
the parameter and $\ell,\la$ refer to the ``source'' and the
``target'' representations respectively.

Here we are interested in $F_{2,0} (t)$ and in $F_{2,2} (t)$,
providing respectively the number of invariants and of covariants
for the $\ell=2$ representation at different orders. These turn
out to be
$$
F_{2,0} \ = \ \frac{1}{(1-t^2) \ (1-t^3)} \ ; \ \
F_{2,2} \ = \ \frac{t+t^2}{(1-t^2) \ (1-t^3)} \ . $$

Expanding these in series up to the order of interest here, we easily obtain
\begin{eqnarray*}
F_{2,0} (t) &=& 1+t^2+t^3+t^4+t^5+2 t^6+t^7+2 t^8+O\left(t^9\right) \ , \\
F_{2,2} (t) &=& t+t^2+t^3+2 t^4+2 t^5+2 t^6+3 t^7+3 t^8+O \left( t^9 \right) \ . \end{eqnarray*}

We refer e.g. to \cite{JB,Sat,Mic,MZZ,Cho,CL} for details on the
Molien function, both in the specific case $G=SO(3)$ and in
general.

\section*{Appendix B. Comparison with previous work}
\label{sec:comparison}

We will now sketch how the same problem could be tackled
following strictly the procedure given in \cite{GL1}. We will just
follow this procedure, referring to \cite{GL1} for its
justification and detail; the understanding of the main body of
the present work does not depend in any way on this appendix.

As mentioned above, the $SO(3)$ representation given by the $J_a$
is not orthogonal w.r.t. the standard scalar product in $\R^5$,
and thus we have to introduce a suitable scalar product for the
procedure described in \cite{GL1} to work; in particular, for the
Sartori $\P$-matrix \cite{AbS,Sar} to be properly defined. This
turns out to be the one associated to the matrix
$$ M \ = \ \begin{pmatrix}  4/3 & 0 & 0 & - 2/3 & 0 \\ 0 & 1 & 0 & 0 & 0 \\
 0 & 0 & 1 & 0 & 0 \\ - 2/3 & 0 & 0 & 4/3 & 0 \\ 0 & 0 & 0 & 0 & 1 \end{pmatrix} \ . $$

The gradients $\nabla T_i (x)$ of the two basic invariants are
easily computed. The $\P$-matrix, defined by $\P_{\a \b} = (
\nabla I_\a , \nabla I_\b )$ with $I_\a$ the invariants, is
$$ \P \ = \ \begin{pmatrix} 4 \, T_2 & 6 \, T_3 \\ 6 \, T_3 & (4/3) \, T_2^2 \end{pmatrix} \ . $$
Proceeding as in \cite{GL1}, the homological operator $\L_0$ is therefore
$$ \L_0 \ = \ 4 \, c_1 \ T_2 \ (\pa / \pa T_2 ) \ . $$
This is discussed in \cite{GL1}; here it suffices to say that
under a change of variables generated by a function  $H_m$
(homogeneous of degree $m+2$), the terms $\Phi_k$ with $m < k$ are
not changed, while the term $\Phi_m$ is changed into $\wt{\Phi}_m
= \Phi_m - \L_0 (H_m)$. (The terms $\Phi_k$ with $k >m$
are changed in a more complex way; this could be described in
precise terms \cite{RNF}, but is not relevant here.)

We should then apply this on homogeneous invariant generating
functions of order up to five or seven (depending if we are in the
framework of section \ref{sec:red6} or \ref{sec:red8}), i.e.
$$ \begin{array}{lll}
H_1 \ = \  k_1 \, T_3 \ , & \ H_2 \ = \ k_2 \, T_2^2 \ , & \
H_3 \ = \ k_3 \, T_2 \, T_3 \ , \\
H_4 \ = \ k_4 \, T_2^3 \ + \ k_5 \, T_3^2 \ , & \ H_5 \ = \ k_6 \,
T_2^2 \, T_3 \ . & \end{array} $$ This yields immediately
$$ \begin{array}{lll}
\L_0 (H_1) \ = \ 0 \ , &
\L_0 (H_2) \ = \ 8 \, c_1 \, k_2 \, T_2^2 \ = \ 8 \, c_1 \, H_2 \ , &
\L_0 (H_3) \ = \ 4 \, c_1 \, k_3 \, T_2 \, T_3 \ = \ 4 \, c_1 \, H_3 \ , \\
\L_0 (H_4) \ = \ 12 \, c_1 \, k_4 \, T_2^3 \ , &
\L_0 (H_5) \ = \ 8 \, c_1 \, k_6 \, T_2^2 \, T_3 \ = \ 8 \, c_1 \, H_5 \ . &
\end{array} $$ Thus the terms $\Phi_2,\Phi_3,\Phi_5$ and the term
proportional to $T_2^3$ in $\Phi_4$, see \eqref{eq:landauT2}, are
in the range of $\L_0$ and by the discussion of \cite{GL1} can be
eliminated from the LdG functional.

Note that, on the other hand, the general results of \cite{GL1} do
not imply the elimination of $\Phi_1$ and of the other term in
$\Phi_4$; this point will be  discussed in a moment.

Moreover, as discussed in \cite{GL1}, while we can conclude that
in the first effective step (assuming we just choose $k_1 = 0$,
i.e. do not perform any change with cubic generating function) one
should choose $k_2 = c_2/(8 c_1)$, the actual determination of
generating functions at higher orders, hence of $k_i$ with $i >
2$, requires to consider in detail the effect of the previous
transformation(s) on the coefficient $c_i$.

The procedure discussed in the present paper does instead provide
a simultaneous computation of the modified coefficients in the
potential and of those identifying the generating function; we
recall this is possible because we know {\it apriori} at which
order we want to stop, while the procedure given in \cite{GL1} can
in principles be pursued up to any order.

Finally, let us come back to the terms which are eliminated in the
present approach but seemingly not in the general one discussed in
\cite{GL1}. It was mentioned in there that the method only
considered first order effects, and that one could take into
account also higher order ones, mimicking what is done in
dynamical systems \cite{CGs,Gae02,RNF}, thus obtaining a ``further
reduction''. The procedure proposed here does take these higher
order effects into account, and hence obtains the further
reduction, avoiding at the same time the relatively sophisticated
mathematical tools which would be needed to obtain a comprehensive
theory (valid at all orders) of it \cite{RNF}.

\section*{Appendix C. Reduction of terms of order seven and eight.}

Here we give explicit formulas for the reduction of terms of order
seven and eight, considered in Section \ref{sec:red8}.

\subsection*{Terms of order seven}

As mentioned in the text, at order seven we have only one invariant,
\beq \^\Phi_5 \ = \ \ga \ T_2^2 \ T_3 \ ; \eeq
by explicit computations it results that
\begin{eqnarray*}
\ga &=& c_7 + 27 c_5 k_1 + 4 c_6 k_1 + 30 c_4 k_1^2 + 54 c_3 k_1^3 +
5 c_4 k^2  + 54 c_3 k_1 k_2 + 9 c_2 k_1^2 k_2 + 3 c_2 k_2^2 +
4 c_3 k_3 + 4 c_2 k_1 k_3 \\
& & + 2 c_1 k_2 k_3 + 18 c_3 k_4 + 18 c_2 k_1 k_4 + 9 c_1 k_2 k_4 +
3 c_2 k_5  + 9 c_1 k_1 k_5 + 2 c_2 k_6 + 6 c_1 k_1 k_6 +
 2 c_1 k_7 + 9 c_1 k_8 \ . \end{eqnarray*}

Setting $k_1,..., k_6$ as determined at lower orders we get
\begin{eqnarray}
\ga &=& [ 112 c_2^5 - 2160 c_1 c_2^3 c_3 + 1080 c_1^2 c_2^2 c_4
 - 54 c_1^2 c_2 (-135 c_3^2 + 54 c_1 c_5 + 8 c_1 c_6) \\
& & + 729 c_1^3 (-4 c_3 c_4 + c_1 (c_7 + 2 c_1 k_7 + 9 c_1 k_8))]
\, / \, [729 c_1^4] \ . \nonumber \end{eqnarray} Note this depends
on the two parameters $k_7$ and $k_8$; thus one of these will
remain undetermined (and can e.g. be set to zero, or to some other
convenient value).

If we choose e.g.
\begin{eqnarray} k_7 &=& [-112 c_2^5 + 2160 c_1 c_2^3 c_3 - 1080 c_1^2 c_2^2 c_4
 + 54 c_1^2 c_2 (-135 c_3^2 + 54 c_1 c_5 + 8 c_1 c_6) \\
& & - 729 c_1^3 (-4 c_3 c_4 + c_1 (c_7 + 9 c_1 k_8))] \
[1458 \, c_1^5]^{-1} \nonumber \end{eqnarray}
then $\ga$ vanishes, and $\^\Phi_5$ with it.

\subsection*{Terms of order eight}

We will refer to the formulas and notation of Sect.\ref{sec:red8ord8}.

In general terms, with $k_1,...,k_7$ taking the values determined
at lower orders (apart from the undetermined $k_4$), we have
{\small
\begin{eqnarray*}
\xi &=& \[ 3609 c_1 c_2^4 c_3  + 27 c_1^2 c_2^2 (516 c_1 c_5 + 32 c_1 c_6 - 1149 c_3^2 )
 + 24 c_1^2 c_2^3 (342 c_1 k_4 - 85 c_4 ) \right. \\
& & \left. + 1944 c_1^3 c_2 (7 c_3 c_4 - 2 c_1 c_7  - 48 c_1 c_3 k_4 + 12 c_1^2 k_8)
 + 2187 c_1^3 (33 c_3^3 - 36 c_1 c_3 c_5 +  8 c_1^2 (c_9 + 2 c_4 k_4 + 3 c_1 k_4^2 \right. \\
& & \left. + 2 c_1 k_9)) - 101 c_2^6  \] \ \[ 17496 c1^5\]^{-1} \ , \\
\eta &=& \[ 5886 c_1 c_2^4 c_3 - 266 c_2^6  + 243 c_1^2 c_2^2 (28 c_1 c_5 + 6 c_1 c_6 - 113 c_3^2 )
 - 18 c_1^2 c_2^3 (172 c_4 + 171 c_1 k_4) \right. \\
& & \left. + 243 c_1^4 (-9 c_4^2 - 18 c_3 c_6 - 54 c_1 c_4 k_4  + c_1 (4 c_8 - 81 c_1 k_4^2))
 + 1458 c_1^3 c_2 (c_3 (13 c_4 + 24 c_1 k_4)  - 2 c_1 (c_7 + 3 c_1 k_8)) \] \\
 & & \ \times \ \[ 971 c_1^5 \]^{-1} \ . \end{eqnarray*}
}

The requirements $\xi = 1$, $\eta = 0$ are satisfied by setting
{\small
\begin{eqnarray}
k_8 &=& \[ 5886 c_1 c_2^4 c_3 - c_1^2 c_2^2  (27459 c_3^2 + 3096 c_2 c_4)
 + c_1^3 c_2 (18954 c_3 c_4 + 6804 c_2 c_5 + 1458 c_2 c_6 - 3078 c_2^2 k_4) \right. \nonumber \\
& & \left. - c_1^4 (2187 c_4^2 + 4374 c_3 c_6 + 2916 c_2 c_7 - 34992 c_2 c_3 k_4)
 + c_1^5 (972 c_8 - 13122 c_4 k_4) - 19683 c_1^6 k_4^2 - 266 c_2^6   \] \nonumber \\
& & \times \ \[ 8748 c_1^5 c_2  \]^{-1}    \ ; \label{eq:k8} \\
k_9 &=&  \[- 57915 c_1 c_2^4 c_3 + c_1^2 c_2^2 (312741 c_3^2 + 30888 c_2 c_4)
 - c_1^3 (216513 c_3^3 + 192456 c_2 c_3 c_4 + 96228 c_2^2 c_5 + 14256 c_2^2 c_6) \right.\nonumber  \\
& & \left.  + c_1^4 (17496 c_4^2 + 236196 c_3 c_5  + 34992 (c_3 c_6 + c_2 c_7))
 + c_1^5 (52488 (1-c_9) - 7776 c_8) + 2431 c_2^6  \] \nonumber \\
 & & \times \ \[ 104976 c_1^6 \]^{-1}   \ . \label{eq:k9} \end{eqnarray}
}

As remarked in Sect.\ref{sec:red8ord8}, these see the appearance
of $c_2$ factors in the denominator, so are valid under the
assumption -- beside that $|c_1|$ is large enough -- that $|c_2|$
is large enough as well. Note that even if we require only
$\eta=0$, we would however get denominators depending on $c_2$ as
well.

On the other hand, the condition $\xi = 1$ (without requiring
$\eta = 0$) can be satisfied by choosing {\small
\begin{eqnarray*} k_9 &=& \[ 101 c_2^6 - 3609 c_1 c_2^4 c_3 +
3 c_1^2 c_2^2 (10341 c_3^2 + 680 c_2 c_4) +
    972 c_1^4 (81 c_3 c_5 + 4 c_2 c_7 + 96 c_2 c_3 k_4) \right. \\
     & & \left. -
    27 c_1^3 (2673 c_3^3 + 504 c_2 c_3 c_4 + 4 c_2^2 (129 c_5 + 8 c_6 + 76 c_2 k_4))  -
    52488 c_1^6 k_4^2 - 5832 c_1^5 (-3 + 3 c_9 + 6 c_4 k_4 + 4 c_2 k_8) \] \\
     & & \times \ [34992 \ c_1^6]^{-1} \ . \end{eqnarray*}
}
In this case, we obtain (as usual, these formulas would be
slightly simplified by a suitable choice for the undetermined
parameters $k_4$ and $k_8$) {\small
\begin{eqnarray*}  \eta &=& \[-266 c_2^6 + 5886 c_1 c_2^4 c_3 +
243 c_1^2 c_2^2 (-113 c_3^2 + 28 c_1 c_5 +  6 c_1 c_6) -
    18 c_1^2 c_2^3 (172 c_4 + 171 c_1 k_4) \right. \\ & & \left. +
    243 c_1^4 (-9 c_4^2 - 18 c_3 c_6 - 54 c_1 c_4 k_4  + c_1 (4 c_8 - 81 c_1 k_4^2))
     - 1458 c_1^3 c_2 (-c_3 (13 c_4 + 24 c_1 k_4) \right.  \\
 & & \left. + 2 c_1 (c_7 + 3 c_1 k_8))\] \ \times \ [972 \ c_1^5]^{-1} \ . \end{eqnarray*}
}

\section*{Appendix D. The transformation for the LdG
potential of degree eight near the main transition point.}

In this Appendix we provide explicit expressions for the
coefficients $k_i$ appearing in $\hb$, see \eqref{eq:covh2}, used
to obtain the simplified potentials \eqref{eq:RP8AB}. For the more
involved expressions we will just give the series expansion in
$c_1$. Higher order coefficients appearing in \eqref{eq:covh2},
i.e. $k_9,k_{10},...$ have no role in this computation and can be
set to zero.

The potential of case $(a)$ is obtained by choosing
\begin{eqnarray*}
k_2 &=& \frac{2 c_{1} c_{6}-3 c_{2} c_{4}}{9
c_{2}^2} \ ; \\
k_3 &=& -\frac{c_{6}}{3 c_{2}} \ ; \\
k_4 &=& 0 \ ; \\
k_5 &=& \frac{4 c_{4}^2+4 c_{3} c_{6}-3 c_{2} c_{7}}{9
   c_{2}^2} \\
   & & + \frac{\left(3 (4 c_{8}+27 c_{9}-4)
   c_{2}^2-2 (81 c_{4} c_{5}+26 c_{4} c_{6}+54
   c_{3} c_{7}) c_{2}+18 c_{3} \left(11
   c_{4}^2+8 c_{3} c_{6}\right)\right)}{54
   c_{2}^4} \,  c_{1} \ + \ O\left(c_{1}^2\right) \ ; \\
k_6 &=& 0 \ ; \\
k_7 &=& -\frac{(c_{8}-1) c_{2}^2-3 c_{4} c_{6}
   c_{2}+c_{1} c_{6}^2}{3 c_{2}^3} \ ; \\
k_8 &=& -\frac{9 c_{9} c_{2}^2-18 c_{4} c_{5} c_{2}-12
   c_{3} c_{7} c_{2}+22 c_{3} c_{4}^2+16
   c_{3}^2 c_{6}}{18 c_{2}^3} \\
   & & \ + \ \frac{\left(-3 (6
   c_{5} c_{6}+c_{4} c_{7}+c_{3} (4
   c_{8}+27 c_{9}-4)) c_{2}^2+2 \left(2
   c_{4}^3+c_{3} (81 c_{5}+34 c_{6}) c_{4}+54
   c_{3}^2 c_{7}\right) c_{2}-18 c_{3}^2 \left(11
   c_{4}^2+8 c_{3} c_{6}\right)\right) }{27
   c_{2}^5} \, c_{1} \\
   & & \ + \ O\left(c_{1}^2\right) \ . \end{eqnarray*}

As for the potential of case $(b)$, this is obtained by choosing
\begin{eqnarray*}
k_2 &=& -\frac{c_{4}}{3 c_{2}} \ ; \\
k_3 &=& \frac{9 c_{5} c_{2}^2-12 c_{3} c_{4}
   c_{2}+c_{1} c_{4}^2}{4 c_{2}^3} \ ; \\
k_4 &=& -\frac{9 c_{5} c_{2}^2-12 c_{3} c_{4}
   c_{2}+c_{1} c_{4}^2}{18 c_{2}^3} \ ; \\
k_5 &=& 0 \ ; \\
k_6 &=& \frac{-6 c_{7} c_{2}^3+8 c_{4}^2 c_{2}^2+c_{1}
   \left((4 c_{8}+27 c_{9}-27) c_{2}^2-2 c_{4} (27
   c_{5}+4 c_{6}) c_{2}+18 c_{3}
   c_{4}^2\right)}{12 c_{2} \left(c_{2}^3-6 c_{1}
   c_{3} c_{2}+c_{1}^2 c_{4}\right)}  \ ; \\
k_7 &=& \frac{c_{2} (-9 c_{4} c_{5}+8 c_{4} c_{6}-4
   c_{2} c_{8})-36 c_{3} \left(c_{4}^2-c_{2}
   c_{7}\right)}{12 c_{2}^3} \\
   & & \ + \ \frac{\left(-3 \left(-27
   c_{5}^2+8 c_{4} c_{7}+8 c_{3} (4 c_{8}+27
   c_{9}-27)\right) c_{2}^2+4 \left(7 c_{4}^3+6
   c_{3} (45 c_{5}+8 c_{6}) c_{4}+216 c_{3}^2
   c_{7}\right) c_{2}-1440 c_{3}^2 c_{4}^2\right)
   }{48 c_{2}^5} \, c_{1} \\
   & & \ + \ O\left(c_{1}^2\right) \ ; \\
k_8 &=& \frac{-9 (c_{9}-1) c_{2}^2+21 c_{4} c_{5}
   c_{2}-10 c_{3} c_{4}^2}{18
   c_{2}^3}+\frac{\left(-144 c_{3}^2 c_{4}^2+216
   c_{2} c_{3} c_{5} c_{4}+c_{2} \left(4
   c_{4}^3-81 c_{2} c_{5}^2\right)\right)
   c_{1}}{216 c_{2}^5} \ + \ O\left(c_{1}^2\right) \ . \end{eqnarray*}

\section*{Acknowledgments}
This work was triggered by participation in the Newton Institute
workshop on ``Symmetry, bifurcation and order parameters''. I
thank the organizers of the workshop, in particular David
Chillingworth, for their invitation, and several participants for
questions and remarks. The first version of this paper was written
while visiting LPTMC (Paris-Jussieu); I thank Maria Barbi for her
hospitality in this occasion. {I would like to warmly thank an
unknown referee for insisting on investigating the application of
the method presented here to the transition region and to the
problem of biaxial phases.} My research is partially supported by
MIUR-PRIN program under project 2010-JJ4KPA.

\end{document}